\documentclass[journal]{IEEEtran}
\usepackage{authblk}
\usepackage{graphicx}
\usepackage{times}
\usepackage{cite}
\usepackage{amssymb}
\usepackage{amsmath,epsfig}
\usepackage[font=footnotesize,labelfont=bf]{caption}
\usepackage{subcaption}
\usepackage{float}
\usepackage{siunitx}
\usepackage{setspace}

\usepackage{indentfirst}
\usepackage{algorithm}
\usepackage{algpseudocode}
\usepackage{bm}
\usepackage{graphicx,xcolor}
\usepackage{color}

\graphicspath{ {Images/} }
\renewcommand{\vec}[1]{\mathbf{#1}}

\DeclareMathAlphabet{\mathpzc}{OT1}{pzc}{m}{it}

\DeclareMathOperator*{\argmin}{argmin}
\makeatletter
\newcommand*{\rom}[1]{\expandafter\@slowromancap\romannumeral #1@}
\makeatother
\usepackage{float}
\usepackage{hyperref}
 \usepackage{lipsum}
\usepackage{cuted}
\begin{document}
\thispagestyle{empty}
\onecolumn
\begin{Large}IEEE Copyright Notice:\end{Large}

\vspace{1cm}
\textcopyright \ 2019 IEEE. Personal use of this material is permitted. Permission from IEEE must be obtained for all other uses, in any current or future media, including reprinting/republishing this material for advertising or promotional purposes, creating new collective works, for resale or redistribution to servers or lists, or reuse of any copyrighted component of this work in other works.

\vspace{1cm}
This article has been accepted for publication in a future issue of IEEE Transactions on Medical Imaging, but has not been fully edited. Content may change prior to final publication. Citation information:
DOI 10.1109/TMI.2019.2911211, IEEE Transactions on Medical Imaging.
URL: \url{http://ieeexplore.ieee.org/stamp/stamp.jsp?tp=&arnumber=8691467&isnumber=4359023}
\newpage

\twocolumn
\setcounter{page}{1}
\title{Approximating the Ideal Observer and Hotelling Observer for binary signal detection tasks by use of supervised learning methods\vspace{-0.5ex}}
\author[1]{Weimin Zhou}
\author[2]{Hua Li}
\author[3]{Mark A. Anastasio\vspace{-2ex}}
\affil[1]{Department of Electrical and Systems Engineering, Washington University in St. Louis, \break St. Louis, MO, 63130 USA (Email: wzhou24@wustl.edu)}
\affil[2]{Department of Bioengineering, University of Illinois at Urbana-Champaign, and the Carle Cancer Center, \break Carle Foundation Hospital,
Urbana, IL, 61801 USA (Email: huali19@illinois.edu, hua.li@carle.com)}
\affil[3]{ Department of Bioengineering, 
University of Illinois at Urbana-Champaign, Urbana, IL, 61801 USA \break(Email: maa@illinois.edu)\vspace{-3ex}}
\maketitle
\begin{abstract}
It is widely accepted that optimization of medical imaging 
system performance should be guided by task-based measures of image quality (IQ).
Task-based measures of IQ quantify the ability of an observer to perform a specific task
such as detection or estimation of a signal (e.g., a tumor).
For binary signal detection tasks, the Bayesian Ideal Observer (IO) sets an upper limit of observer performance and has
been advocated for use in optimizing medical imaging systems and data-acquisition designs.
Except in special cases, determination of the IO test statistic is
 analytically intractable. Markov-chain Monte Carlo (MCMC) techniques can be employed to approximate 
 IO detection performance, but their reported applications have been limited to relatively simple object models.
In cases where the IO test statistic is difficult to compute, 
the Hotelling Observer (HO)
can be employed.  
To compute the HO test statistic, potentially large covariance matrices
 must be accurately estimated and subsequently inverted,  
 which can present computational challenges.
This work investigates supervised learning-based  methodologies for approximating the IO and HO test statistics.
Convolutional neural networks (CNNs) and single-layer neural networks (SLNNs)
 are employed to approximate the IO and HO test statistics, respectively.
Numerical simulations were conducted for both \textit{signal-known-exactly} (SKE) and \textit{signal-known-statistically} (SKS) signal detection tasks.
The considered background models include the lumpy object model and the clustered lumpy object model.
The measurement noise models considered are Gaussian, Laplacian, and mixed Poisson-Gaussian.
The performances of the supervised learning methods are assessed via receiver operating 
characteristic (ROC) analysis and the results are compared to those produced
 by use of traditional numerical methods or analytical calculations when feasible.
The potential advantages of the proposed supervised learning approaches for approximating
the IO and HO test statistics are discussed. 

\end{abstract}


%
\section{Introduction}
 Medical imaging systems commonly are assessed, validated, and optimized using task-specific measures of image quality that quantify the ability of an observer to perform a specific task~\cite{barrett2013foundations, kupinski2003ideal, park2007channelized, park2009efficient, shen2006using}.
 When optimizing imaging systems for signal detection tasks (e.g., detection of a tumor),
it has been  advocated to use the performance of the Bayesian Ideal Observer (IO) as a figure-of-merit (FOM).
In this way, the imaging system can be optimized in such a way that the amount of task-specific
information in the measurement data is maximized.
The IO for a binary signal detection task implements a test statistic given by the likelihood ratio and maximizes the area under the receiver operating characteristic (ROC) curve~\cite{wagner1985unified}. 
 The IO has also been employed to assess the efficiency of human observers on signal detection tasks~\cite{liu1995object}.

 The IO test statistic is generally  a non-linear function of the image data and,
 except in some special cases,  cannot be determined analytically.
Because of this,  sampling-based methods that employ Markov-chain Monte Carlo (MCMC) techniques
have been developed to approximate the IO test statistic for medical imaging applications~\cite{kupinski2003ideal, abbey2008ideal}. 
 However, current applications of these methods have been
 limited to relatively simple object models that include parameterized torso phantoms~\cite{he2008toward},
  lumpy background models~\cite{kupinski2003ideal}, and a binary texture model~\cite{abbey2008ideal}.
 To the best of our knowledge, applications of MCMC methods to approximate the IO test statistic 
for more sophisticated object models---such as the clustered lumpy background (CLB) model that has been used to synthesize mammographic images---have not been reported to date.

 When the IO is intractable, the Hotelling Observer (HO) can be employed
to optimize imaging systems for signal detection tasks~\cite{barrett2015task,reiser2010task,sanchez2014task,glick2002investigation}.
  The HO employs the Hotelling discriminant, which is the population equivalent of the Fisher linear discriminant~\cite{barrett2013foundations}, and is optimal among all linear observers in the sense that it maximizes the signal-to-noise ratio of the test statistic~\cite{barrett2013foundations,barrett1992linear,barrett1993model}.  
 However, implementation of the HO is also not without challenges. Specifically,
it requires the estimation and inversion of a covariance matrix that can be
 enormous~\cite{barrett2001megalopinakophobia}.
Different strategies for circumventing this difficulty exist~\cite{barrett2015task}.
For use in detection tasks where background variability is considered and the measurement noise covariance matrix is known, methods for the estimation and inversion of these large
 covariance matrices by use of a covariance matrix decomposition are available~\cite{barrett2013foundations}.
It has been demonstrated, however, that in certain situations the
 use of the covariance decomposition can result in a significant
bias in the HO performance~\cite{kupinski2007bias}.
  Alternatively, to avoid an explicit inversion
of the covariance matrix, an iterative algorithm can be employed to estimate the
Hotelling test statistic~\cite{barrett2013foundations}.
Finally, a variety of channelized HOs that utilize efficient channels have
been proposed for approximating the HO in a computationally tractable
 way~\cite{barrett1998stabilized,gallas2003validating,park2009efficient}.

Supervised learning-based approaches hold significant promise for the design and
 implementation of model observers for optimizing imaging systems~\cite{brankov2009learning,wernick2010machine, massanes2017evaluation,alnowami2018deep}.
Recent efforts have primarily focused on training anthropomorphic model observers using
deep learning~\cite{massanes2017evaluation,kopp2018evaluation,kopp2018cnn}. 
The extent to which deep learning-based methods can benefit such
 applications remains a topic of investigation due to the difficulty of acquiring large amounts of labeled data in medical
imaging applications.
When optimizing imaging systems and data-acquisition designs,  computer-simulated data
can sometimes be employed~\cite{kupinski2003ideal}.
In such applications, large amounts of labeled data can be generated and it can be feasible to train complicated inference models to be employed as
 model observers for assessing task-based measures of image quality.

Artificial neural networks (ANNs) with sufficiently
complex architectures are known to be able to approximate any continuous function~\cite{hornik1989multilayer}. 
Accordingly, in principle, ANNs can be trained to approximate functions that represent test statistics of model observers.
For example,  Kupinski~\emph{et al.} investigated the use of fully-connected neural networks (FCNNs) to approximate the test
statistic of an IO that acted on low-dimensional vectors of extracted image features~\cite{kupinski2001ideal}.
More recently, Zhou and Anastasio  employed convolutional neural networks (CNNs)
to approximate the IO test statistic that acted directly on images for a simple signal-known-exactly and background-known-exactly (SKE/BKE) binary signal detection task, 
and demonstrated the use of modern deep learning technologies
for approximating  IOs~\cite{zhou2018learning}.

In this work, supervised learning-based methods that employ ANNs for approximating
  the IO test statistic are explored systematically for binary signal detection tasks
 in which the observer acts on 2D image data. The detection tasks considered are of varying
difficulty, and address both background and signal randomness in combination with
 different measurement noise models.
In order to approximate the generally nonlinear IO test statistic, CNNs are employed.
For the special case of the HO, an alternative supervised learning methodology
is proposed that employs single-layer neural networks (SLNNs) for learning the Hotelling
 template without the need for explicitly estimating and inverting covariance matrices.
The signal detection performance is assessed via receiver operating characteristic (ROC) analysis~\cite{metz1986roc, barrett2013foundations}. The results produced by the proposed supervised learning methods are compared to those produced by use of traditional numerical methods or analytical calculations when feasible.
 The potential advantages of the proposed supervised learning approaches for approximating the IO and HO test statistics are discussed.

The remainder of this article is organized as follows. 
In Sec.~\ref{sec:background}, the salient aspects of binary
 signal detection theory are reviewed and previous works on approximating the IO test statistic by use of ANNs are summarized.
A novel methodology that employs SLNNs to approximate the HO test statistic is developed in  Sec.~\ref{sec:SLNN}.
The numerical studies and results of the proposed methods for approximating the IO and HO for signal detection tasks with different object models and noise models are provided in Sec.~\ref{sec:num} and Sec.~\ref{sec:results}.
Finally, the article concludes with a discussion of the work in Sec.~\ref{sec:concludes}.
 {\vspace{-0.42cm}}
 \section{Background}
\label{sec:background}
 \vspace{-0.05cm}
Consider a linear digital imaging system that is described~as:
\vspace{-0.2cm}
\begin{equation}
\vec{g} = \mathcal{H}{f}(\vec{r}) + \vec{n}, 
\end{equation}
where $\vec{g}\in\mathbb{R}^{M\times 1}$ is a vector that describes the measured image data,
 ${f}(\vec{r})$ is the object function with a spatial coordinate $\vec{r}\in \mathbb{R}^{k\times 1}$, $k=2$ or $3$,
 $\mathcal{H}$ denotes a continuous-to-discrete (C-D) imaging operator that maps $\mathbb{L}_2(\mathbb{R}^{k})\rightarrow\mathbb{R}^{M\times 1}$, and $\vec{n}\in\mathbb{R}^{M\times 1}$ is the measurement noise.
Because $\vec{n}$ is a random vector, so is the measured image data $\vec{g}$.
Below, the object function ${f}(\vec r)$ will be viewed as being either deterministic or stochastic,
 depending on the specification of the signal detection task.  When its spatial dependence
is not important to highlight, the notation $\vec f$ will be
employed to denote $f(\vec r)$.  The same notation will be employed with other functions.
 \vspace{-0.22cm}
\subsection{Formulation of binary signal detection tasks}
\vspace{-0.08cm}
A binary signal detection task requires an observer to classify 
an image as satisfying either a signal-present hypothesis ($H_1$) or a signal-absent hypothesis ($H_0$). 
The imaging processes under these two hypotheses can be described as:
\begin{subequations}\label{eq:imgH}
\begin{equation}
H_{0}: \vec{g} = \mathcal{H}\vec{f}_{b} + \vec{n} \equiv \vec{b} + \vec{n}, 
\end{equation}
\vspace{-0.5cm}
\begin{equation}
H_{1}:  \vec{g} = \mathcal{H}(\vec{f}_{b}+\vec{f}_{s}) + \vec{n} \equiv \vec{b} + \vec{s} + \vec{n},
\end{equation}
\end{subequations}
where $\vec{f}_{b}$ and $\vec{f}_{s}$ represent the background and signal {functions}, respectively, 
$\vec{b}\equiv\mathcal{H}\vec{f}_{b}$ is the background image and $\vec{s}\equiv\mathcal{H}\vec{f}_{s}$ is the signal image.  In a signal-known-exactly (SKE) detection task, $\vec{f}_{s}$ is non-random, whereas in
a signal-known-statistically (SKS) detection task it is a random process.
Similarly, in a background-known-exactly (BKE) detection task, $\vec{f}_{b}$ is non-random, whereas in
a background-known-statistically (BKS) detection task it is a random process.
Let ${b}_m$ and ${s}_m$ denote the $m^{th}$ $(1\leq m \leq M)$ component of $\vec{b}$ and $\vec{s}$, respectively.
When $\mathcal{H}$ is a linear operator, 
as in the numerical studies presented later,
these quantities are defined as:
\begin{subequations}
\label{eq:imgh}
\begin{equation} \label{eq:bb}
{b}_m = \int_{\mathbb{R}^k} d\vec{r}\;{{h}_m (\vec{r}) {f_b}(\vec{r}) },
\end{equation}
\vspace{-0.1cm}
\begin{equation}\label{eq:ss}
{s}_m = \int_{\mathbb{R}^k}  d\vec{r}\;{{h}_m (\vec{r}) {f_s}(\vec{r}) },
\end{equation}
\end{subequations}
where ${h}_m({\vec{r}})$ is the point response
 function of the imaging system associated with the $m^{th}$ measurement~\cite{barrett2013foundations}.

To perform a binary signal detection task, an observer computes a test
statistic $t(\vec{g})$ that maps the measured image $\vec{g}$ to a real-valued scalar variable,
which is compared to a predetermined threshold $\tau$ to classify $\vec{g}$ as satisfying $H_0$ or $H_1$.
 By varying the threshold $\tau$, a ROC curve can be plotted to depict the trade-off between the false-positive fraction (FPF) and the true-positive fraction (TPF)~\cite{metz1986roc, barrett2013foundations}. The area under the ROC curve (AUC) can be subsequently calculated to quantify the signal detection performance.
 \vspace{-0.22cm}
\subsection{Bayesian Ideal Observer and Hotelling Observer}
\label{subsec:IO}
\vspace{-0.08cm}
Among all observers, the IO sets an upper performance limit for binary signal detection tasks. 
The IO test statistic is defined as any monotonic transformation of the likelihood ratio $\Lambda_\text{LR}(\vec{g})$, 
which is defined as~\cite{barrett2013foundations, kupinski2003ideal, kupinski2001ideal}:
\begin{equation}
\Lambda_\text{LR}(\vec{g}) = \frac{p(\vec{g}|H_1)}{p(\vec{g}|H_0)}.
\end{equation}
Here, $p(\vec{g}|H_0)$ and $p(\vec{g}|H_1)$ are conditional probability density functions
 that describe the measured data $\vec{g}$ under hypothesis $H_0$ and $H_1$, respectively. 
It will prove useful to note that one monotonic transformation of $\Lambda_\text{LR}(\vec{g})$ is the
 posterior probability $\Pr(H_1 | \vec{g})$: 
 \vspace{-0.1cm}
\begin{equation} \label{eq:pos}
\Pr(H_1|\vec{g}) = \frac{\left[\Pr(H_1)/\Pr(H_0)\right] \Lambda_\text{LR}(\vec{g})}{1+\left[\Pr(H_1)/\Pr(H_0)\right] \Lambda_\text{LR}(\vec{g})},
\end{equation}  
where $\Pr(H_0)$ and $\Pr(H_1)$ are the prior probabilities associated with the two hypotheses.

When the IO test statistic cannot be determined analytically, 
the HO is sometimes employed to assess task-based measures of image quality.
The HO employs the Hotelling discriminant that is the population equivalent of the Fisher linear discriminant~\cite{barrett2013foundations}.
The HO test statistic $t_{\text{HO}}(\vec{g})$ is computed as:
\vspace{-0.1cm}
\begin{equation}\label{eq:HO_cal}
t_\text{HO}(\vec{g}) =  \vec{w}_\text{HO}^T \vec{g},
\end{equation}
where $\vec{w}_\text{HO}\in\mathbb{R}^{M\times 1}$ is the Hotelling template.
Let $\bar{\vec{g}}(\vec{f})\equiv\langle \vec{g} \rangle_{\vec{g}|\vec{f}}$
denote the conditional mean of the image data given an object function.
Similarly, let $\bar{\bar{\vec{g} }}_j\equiv \langle \bar{\vec{g}}(\vec{f}) \rangle_{\vec{f}|H_j}$
denote the conditional mean averaged with respect to object randomness associated with $H_j$ ($j=0, 1$).
The Hotelling template $\vec{w}_\text{HO}$ is defined as~\cite{barrett2013foundations}:
\vspace{-0.26cm} 
\begin{equation} \label{eq:w_HO}
\vec{w}_\text{HO}=\left[\frac{1}{2}(\vec{K}_0 + \vec{K}_1)\right]^{-1}\Delta\bar{\bar{\vec{g}}}.
\end{equation}
Here,  $\vec{K}_j =\big\langle \langle [\vec{g} -\bar{\bar{\vec{g} }}_j][\vec{g} -\bar{\bar{\vec{g} }}_j]^{T}  \rangle_{\vec{g}|\vec{f}} \big\rangle_{\vec{f}|H_j}$  is the covariance matrix of the measured data $\vec{g}$ under the hypothesis $H_j$ ($j=0, 1$), and $\Delta \bar{\bar{\vec{g}}} = \bar{\bar{\vec{g} }}_1 - \bar{\bar{\vec{g} }}_0$ is the difference between the mean of the measured data $\vec{g}$ under the two hypotheses. 
It is useful to note that the covariance matrix $\vec{K}_j$ can be decomposed as~\cite{barrett2013foundations}:
\vspace{-0.3cm}
\begin{equation}\label{eq:K}
\begin{split}
\vec{K}_j = &\big\langle  \langle [\vec{g} - \bar{\vec{g}}(\vec{f})] [\vec{g} - \bar{\vec{g}}(\vec{f})]^T   \rangle_{\vec{g}|\vec{f}}  \big\rangle_{\vec{f}|H_j} \\
& + \langle  [\bar{\vec{g}}(\vec{f}) - \bar{\bar{\vec{g}} }_j] [\bar{\vec{g}}(\vec{f}) - \bar{\bar{\vec{g}} }_j]^T  \rangle_{\vec{f}|H_j} \\
\equiv &\langle{\vec{K}}_{\vec{n}|\vec{f}} \rangle_{\vec{f}|H_j}+ \vec{K}_{\bar{\vec{g}}(\vec{f}) | H_j}.
\end{split}
\end{equation}
In Eq. (\ref{eq:K}), the first term $\langle{\vec{K}}_{\vec{n}|\vec{f}} \rangle_{\vec{f}|H_j}$ is the mean of the noise covariance matrix ${\vec{K}}_{\vec{n}|\vec{f}}$ averaged over $\vec{f}$ under the hypothesis $H_j$.
 The second term $\vec{K}_{\bar{\vec{g}}(\vec{f}) | H_j}$ is the covariance matrix associated with the object $\vec{f}$ under the hypothesis $H_j$. 

The signal-to-noise ratio associated with a test statistic $t$, denoted as $\text{SNR}_t$, is defined as:
\begin{equation}\label{eq:SNR2}
\text{SNR}_t = \frac{\langle t\rangle_1 - \langle t\rangle_0}{\sqrt{\frac{1}{2}\sigma_0^2+\frac{1}{2}\sigma_1^2}},
\end{equation}
where $\langle t \rangle_j$ and $\sigma_j^2 = \big\langle(t-\langle t\rangle_j)^2 \big\rangle_j$ are the mean and variance of $t$ under the hypothesis $H_j$ ($j=0, 1$). Similar to the AUC, 
$\text{SNR}_t$ is a commonly employed FOM of signal detectability that can be employed to
guide the optimization of imaging systems. Whereas the IO maximizes the AUC among all observers,
the HO maximizes the value of $\text{SNR}_t$ among all linear observers that can be computed as~\cite{barrett2013foundations,barrett1993model}:
 \vspace{-0.1cm}
\begin{equation}\label{eq:SNR_HO}
\text{SNR}_{HO}^2 = \Delta\bar{\bar{\vec{g}}}^T\vec{w}_\text{HO}.
\end{equation}

\subsection{Previous works on approximating the IO test statistic by use of ANNs}
A feed-forward ANN is a system of connected artificial neurons that are computational units described by adjustable real-valued parameters called weights~\cite{schmidhuber2015deep, lecun2015deep}. 
A sufficiently complex ANN possesses the ability to approximate any continuous function~\cite{hornik1989multilayer}. Accordingly, ANNs can be trained to approximate functions that represent test statistics of model observers.
Previous published results indicate the feasibility of using ANNs to approximate IOs~\cite{kupinski2001ideal, zhou2018learning}.
For example, Kupinski~\emph{et al.}~\cite{kupinski2001ideal} applied fully-connected neural networks (FCNNs), which are a conventional type of feed-forward ANNs, 
to approximate the test statistic for an IO acting on low-dimensional vectors of extracted image features.
It was demonstrated that~\cite{kupinski2001ideal}, given sufficient training data and an 
 ANN of sufficient representation capacity,
 the test statistic of the IO acting on a low-dimensional vector of image features
could be accurately approximated.
 However, ordinary ANNs, such as FCNNs, do not scale well to high-dimensional data (e.g., images) because each neuron in FCNNs is fully connected to all neurons in the previous layer,
which limits the dimension of the input layer and depth of the models that can
be trained effectively. 
As such, FCNNs are not well suited for use as numerical observers that act directly on image data.

Modern deep learning approaches that employ convolutional neural networks~(CNNs) have been developed to address this limitation~\cite{lecun2015deep,lawrence1997face, cirecsan2012multi,garcia2004convolutional}.
A comprehensive review of CNNs for image classifications can be found in~\cite{rawat2017deep}.
Recently, motivated by the success of CNNs in image classification tasks, Zhou and Anastasio~\cite{zhou2018learning} investigated a supervised learning-based method to approximate the test statistic of an IO that acts directly on 2D images by using CNNs. 
The basic idea is to identify a CNN that can approximate $\Pr(H_1\vert \vec g)$ which, as described by  Eq. (\ref{eq:pos}),
is a monotonic transformation of the likelihood ratio.
In that preliminary work, the feasibility of using CNNs to approximate an IO for a simple SKE/BKE object model was explored. 
As an extension of that preliminary study, supervised learning-based methods that employ CNNs and SLNNs for approximating test statistics of the IO and HO acting on 2D measured images with various object and noise models are systematically explored in this work. 
 \vspace{-0.06cm}
\subsection{Maximum likelihood estimation of CNN weights for approximating the IO test statistic}
 To train a CNN for approximating the posterior probability $\Pr(H_1|\vec{g})$, 
 the sigmoid function is employed in the last layer of the CNN; in this way
 the output of the CNN can be interpreted as probability.
Let the set of all weights of neurons in a CNN be denoted by the vector $\bm{\Theta}$ and
 denote the output of the CNN as $\Pr(H_1|\vec{g},\bm{\Theta})$.
It should be noted that the vertical bar in $\Pr(H_1|\vec{g},\bm{\Theta})$ 
has two usages: to denote that the probability of $H_1$ is  conditioned on  $\vec{g}$ 
and to denote that the function is parameterized by the nonrandom weight vector $\bm{\Theta}$.
The goal of training the CNN is to determine a vector $\bm{\Theta}$ such that the
 difference between the CNN-approximated posterior probability $\Pr(H_1|\vec{g},\bm{\Theta})$ and the actual posterior probability $\Pr(H_1|\vec{g})$ is small.
 The posterior $\Pr(H_0|\vec{g})$ can be subsequently approximated by  $\Pr(H_0|\vec{g}, \bm{\Theta})\equiv 1-\Pr(H_1|\vec{g}, \bm{\Theta})$.

A supervised learning-based method can be
 employed to approximate the maximum likelihood (ML) estimate of $\bm{\Theta}$~\cite{kupinski2001ideal}.
Let $y\in \{0,1\}$ denote the image label, where $y=0$ and $y=1$ correspond to the hypothesis $H_0$ and $H_1$, respectively. 
The ML estimate of $\bm{\Theta}$ can be obtained by minimizing the generalization error defined as the ensemble average of cross-entropy over distribution $p(\vec{g}, y)$~\cite{kupinski2003ideal}:
\begin{equation}\label{eq:cr}
\bm{\Theta}_\text{ML} = \argmin_{\bm{\Theta}} \left\langle -\log\big[\Pr(y|\vec{g}, \bm{\Theta})\big]\right\rangle_{(\vec{g}, y)},
\end{equation}
where $\langle . \rangle_{(\vec{g}, y)}$ denotes the mean over the probability density $p(\vec{g}, y)$.
If $\Pr(H_1|\vec{g},\bm{\Theta})$ can represent any functional form, $\Pr(H_1|\vec{g},\bm{\Theta}_\text{ML}) = \Pr(H_1|\vec{g})$ when Eq. (\ref{eq:cr}) is minimized~\cite{kupinski2003ideal}.
To see this, one can rewrite the negative cross-entropy as:
 \vspace{-0.2cm}
\begin{equation}\label{eq: append2}
\begin{split}
 \left\langle \log\big[\Pr(y|\vec{g}, \bm{\Theta})\big]\right\rangle_{(\vec{g}, y)}&=
\int_{\mathbb{R}^M} \Big[\log\big(\Pr(H_1|\vec{g},{\bm{\Theta}})\big)p(\vec{g}, H_1) \\
+\log&\big(1-\Pr(H_1|\vec{g},{\bm{\Theta}})\big)p(\vec{g}, H_0)\Big] d^{M}\vec{g}.
\end{split}
\end{equation}
When the CNN is sufficiently complex to represent any functional form, the task of finding $\bm{\Theta}_\text{ML}$ becomes finding the optimal $\Pr(H_1|\vec{g},{\bm{\Theta}})$ that maximizes Eq. (\ref{eq: append2}). Consider the gradient of Eq. (\ref{eq: append2}) with respect to $\Pr(H_1|\vec{g},{\bm{\Theta}})$:
\begin{equation}\label{eq: append3}
\begin{split}
\nabla&_{\Pr(H_1|\vec{g},{\bm{\Theta}})} \left\langle \log\big[\Pr(y|\vec{g}, \bm{\Theta})\big]\right\rangle_{(\vec{g}, y)}
= \\
&\left[\frac{\Pr(H_1|\vec{g})}{\Pr(H_1|\vec{g}, \bm{\Theta})}
-\frac{1-\Pr(H_1|\vec{g})}{1-\Pr(H_1|\vec{g}, \bm{\Theta})}\right]p(\vec{g}).
\end{split}
\end{equation}
For $\vec{g} \in \{\vec{g}|p(\vec{g})\neq 0\}$, Eq. (\ref{eq: append3}) equals zero only when $\frac{\Pr(H_1|\vec{g})}{\Pr(H_1|\vec{g}, \bm{\Theta})}
=\frac{1-\Pr(H_1|\vec{g})}{1-\Pr(H_1|\vec{g}, \bm{\Theta})}$, from which $\Pr(H_1|\vec{g}, \bm{\Theta}_\text{ML}) = \Pr(H_1|\vec{g})$.

Given a set of independent labeled training data $\left\{(\vec{g}_{i},y_{i})\right\}_{i=1}^{N}$, 
$ \bm{\Theta}_\text{ML}$ can be estimated by minimizing the empirical error, which is the average of the cross-entropy over the training dataset:
\begin{equation} \label{eq:IO_loss}
\hat{\bm{\Theta}}_\text{ML} =\argmin_{\bm{\Theta}} \left[ -\sum_{i=1}^{N}  \log\big(\Pr(y_i|\vec{g}_i, \bm{\Theta})\big)\right],
\end{equation}
where $\hat{\bm{\Theta}}_\text{ML}$ is the empirical estimate of ${\bm{\Theta}}_\text{ML}$.
The IO test statistic is subsequently approximated as $\Pr(H_1|\vec{g}, \hat{\bm{\Theta}}_\text{ML})$. However, if the training dataset is small,
directly minimizing the empirical error can cause overfitting and large generalization errors~\cite{goodfellow2016deep}.
To reduce the rate at which overfitting happens, mini-batch stochastic gradient descent algorithms can be employed~\cite{goodfellow2016deep}. In online learning, these mini-batches are
drawn on-the-fly from the joint distribution $p(\vec{g}, y)$~\cite{goodfellow2016deep}.

\section{Approximation of the HO test statistic by use of SLNNs} \label{sec:SLNN}
Below, a novel supervised learning-based method is proposed for learning the HO test statistic.
\vspace{-0.3cm}
\subsection{Training the HO by use of supervised learning}
As described by Eq. (\ref{eq:HO_cal}),
the HO test statistic is a linear function of the measured image $\vec{g}$. 
{Linear functions can be modeled by a single-layer neural network (SLNN) that possesses only a single fully connected layer. 
Denote the vector of weight parameters in the SLNN as $\vec{w}\in \mathbb{R}^{M\times 1}$. The output of a SLNN can be computed as: }
 \vspace{-0.1cm}
\begin{equation}
t_{\text{SLNN}} (\vec{g})= \vec{w}^T\vec{g}.
\end{equation}
To approximate $t_{\text{HO}}(\vec{g})$ by $t_{\text{SLNN}}(\vec{g})$, 
a SLNN can be trained by maximizing $\text{SNR}_t$ by solving the following optimization problem:
\vspace{-0.11cm}
\begin{equation} \label{eq:opt}
\begin{aligned}
& \underset{\mathbf{w}}{\text{minimize} } 
& & \frac{1}{2}\left \langle [\mathbf{w}^T\mathbf{g}-  \mathbf{w}^T\bar{\bar{\vec{g}}}_0]^2 \right\rangle_0 + \frac{1}{2}\left \langle [\mathbf{w}^T\mathbf{g}-  \mathbf{w}^T\bar{\bar{\vec{g}}}_1]^2 \right\rangle_1\\
& \text{subject to} & &  \mathbf{w}^T\bar{\bar{\vec{g}}}_1 - \mathbf{w}^T\bar{\bar{\vec{g}}}_0=C,
 \end{aligned}
 \end{equation}
 where $C$ is any positive number. 
 The Lagrangian function related to this constrained optimization problem can be computed as: 
 \vspace{-0.11cm}
 \begin{equation}\label{eq:L1}
 \begin{split}
 L(\mathbf{w}, \lambda) =  &\frac{1}{2}\left \langle [\mathbf{w}^T\mathbf{g}-  \mathbf{w}^T\bar{\bar{\vec{g}}}_0]^2 \right\rangle_0 + \frac{1}{2}\left \langle [\mathbf{w}^T\mathbf{g}-  \mathbf{w}^T\bar{\bar{\vec{g}}}_1]^2 \right\rangle_1 \\
& - \lambda ( \mathbf{w}^T\bar{\bar{\vec{g}}}_1 - \mathbf{w}^T\bar{\bar{\vec{g}}}_0-C).
 \end{split}
 \end{equation}
 The optimal solution $\mathbf{w}^*$ satisfies the Lagrange multiplier conditions:
 \begin{subequations} \label{subeq:L}
\label{eq:imgH_s}
 \vspace{-0.1cm}
\begin{equation}
\nabla_{\mathbf{w}}L(\mathbf{w}^*, \lambda^*) =\left[ \vec{K}_0 + \vec{K}_1\right] \mathbf{w}^* -\lambda^*  \Delta \bar{\bar{\vec{g}}} =0, 
\end{equation}
 \vspace{-0.2cm}
\begin{equation}
\nabla_{\lambda}L(\mathbf{w}^*, \lambda^*) = -\left[\mathbf{w}^{*T}  \Delta \bar{\bar{\vec{g}}}  -C\right] = 0, 
\end{equation}
\end{subequations}
 where $\lambda^*$ is the Lagrange multiplier.  According to Eq. (\ref{subeq:L}):
  \begin{subequations} 
\label{eq:imgH_s}
\begin{equation}
\mathbf{w}^*= \left[\frac{1}{\lambda^*}(\vec{K}_0 + \vec{K}_1)\right]^{-1}\Delta \bar{\bar{\vec{g}}},
\end{equation}
 \vspace{-0.2cm}
\begin{equation}
\lambda^* = \frac{C}{ \Delta \bar{\bar{\vec{g}}}^T(\vec{K}_0 + \vec{K}_1)^{-1} \Delta \bar{\bar{\vec{g}}}}.
\end{equation}
\end{subequations}
Because Eq. (\ref{eq:L1}) is convex, $\mathbf{w}^*$ is the global minimum of $L(\mathbf{w}, \lambda^*)$ and the constrained optimization problem defined in Eq. (\ref{eq:opt}) can be solved by minimizing $L(\mathbf{w}, \lambda^*)$ with respect to $\mathbf{w}$, which is equivalent to minimizing $L(\mathbf{w}, \lambda^*) - \lambda^* C$ with respect to $\mathbf{w}$. Hence, the generalization error
 to be minimized is defined as:
 \begin{equation}\label{eq:obj}
 \begin{split}
 &l(\mathbf{w}) \equiv L(\mathbf{w}, \lambda^*) - \lambda^* C \\
 &= \frac{1}{2}\left \langle [\mathbf{w}^T (\mathbf{g}-\bar{\bar{\vec{g}}}_0 )]^2 \right\rangle_{0} + \frac{1}{2}\left \langle [\mathbf{w}^T (\mathbf{g}-\bar{\bar{\vec{g}}}_1)]^2 \right\rangle_{1} 
 - \lambda^* \mathbf{w}^T\Delta \bar{\bar{\vec{g}}}.
 \end{split}
 \end{equation}
 In order to have $\mathbf{w}^*= \mathbf{w}_\text{HO}$, $\lambda^*$ is set to 2. 
 
 Given $N$ labeled image data $\{\mathbf{g}_i, y_i\}_{i=1}^{N}$ in which half of them are signal-absent and the others are signal-present, the empirical error to be minimized is:
\begin{equation} \label{eq:L3}
\begin{split}
   \hat{l}(\mathbf{w}) = 
   \frac{1}{N} \sum_{i=1}^{N}&\Big\{(1-y_i)\left[\mathbf{w}^T (\mathbf{g}_i- \hat{\vec{g}}_0 )\right]^2 \\ 
   + &y_i \left[\mathbf{w}^T (\mathbf{g}_i-\hat{\vec{g}}_1 )\right]^2\Big\} 
 - 2\mathbf{w}^T\Delta \hat{\vec{g}},
 \end{split}
\end{equation}
where $\hat{\vec{g}}_0 =  \frac{2}{N}\sum_{i=1}^{N} (1-y_i)\vec{g}_i$, $\hat{\vec{g}}_1 =  \frac{2}{N}\sum_{i=1}^{N} y_i \vec{g}_i$, and $\Delta\hat{\vec{g}} = \hat{\vec{g}}_1- \hat{\vec{g}}_0$.

Any gradient-based algorithm can be employed to minimize Eq. ($\ref{eq:L3}$) to learn the empirical estimate of the Hotelling template, which is equivalent to the template employed by the Fisher linear discriminant. Because this method does not require estimation and inversion of a covariance matrix, it can scale well to large images.

\vspace{-0.2cm}
\subsection{Training the HO by use of a covariance-matrix decomposition}\label{sec:HO_cmd}
Methods have been developed previously to estimate and invert empirical covariance matrices by use of a covariance-matrix decomposition~\cite{barrett2013foundations, kupinski2007bias}. 
As stated in Eq. (\ref{eq:K}), the covariance matrix $\vec{K}_j$ can be decomposed into the component associated with the object randomness $\vec{K}_{\bar{\vec{g}}(\vec{f}) | H_j}$ and that associated with the noise randomness $\langle{\vec{K}}_{\vec{n}|\vec{f}} \rangle_{\vec{f}|H_j}$. 
To invert the full covariance matrix for computing the HO test statistic, $\langle{\vec{K}}_{\vec{n}|\vec{f}} \rangle_{\vec{f}|H_j}$ is assumed known and $\vec{K}_{\bar{\vec{g}}(\vec{f}) | H_j}$ needs to be estimated from samples of background and signal images.
When uncorrelated noise is considered, $\langle{\vec{K}}_{\vec{n}|\vec{f}} \rangle_{\vec{f}|H_j}$ is a diagonal matrix. For applications where detectors introduce correlations in the measurements, $\langle{\vec{K}}_{\vec{n}|\vec{f}} \rangle_{\vec{f}|H_j}$ is banded and may be a nearly diagonal matrix~\cite{barrett2013foundations}.
In this subsection, an alternative method is provided to approximate the HO test statistic by use of a covariance-matrix decomposition.

According to the covariance-matrix decomposition stated in Eq. (\ref{eq:K}), the variance of the test statistic can be computed as: 
 \begin{equation}
\left \langle (\mathbf{w}^T\mathbf{g}-\langle \mathbf{w}^T\mathbf{g} \rangle_{j})^2 \right\rangle_{j} =  
\vec{w}^T\vec{K}_{\vec{\bar{g}(f)}|H_j}\vec{w} +
\vec{w}^T\langle \vec{K_{n|f}} \rangle_{\vec{f}|H_j}\vec{w}.
 \end{equation}
Denote $\frac{1}{2} (\langle \vec{K_{n|f}} \rangle_{\vec{f}|H_0} + \langle \vec{K_{n|f}} \rangle_{\vec{f}|H_1})$ as $\overline{\vec{K}}_\vec{n}$, which is assumed known. 
 The generalization error 
defined in Eq. (\ref{eq:obj}) can be reformulated as:
 \begin{equation}
 \begin{split}
  l(\mathbf{w}) = &\left \langle (\mathbf{w}^T\mathbf{b}- \mathbf{w}^T \bar{\vec{b}})^2 \right\rangle_{\mathbf{f}_b} + 
   \frac{1}{2}\left \langle (\mathbf{w}^T\mathbf{s}- \mathbf{w}^T \bar{\vec{s}})^2 \right\rangle_{\mathbf{f}_s}  \\
  &+\mathbf{{w}}^T\overline{\vec{K}}_\vec{n}\mathbf{{w}} -
  2 \mathbf{w}^T\bar{\vec{s}},
  \end{split}
  \end{equation}
 where $\bar{\vec{b}} = \langle \vec{b} \rangle_{\vec{f}_b}$, and $\bar{\vec{s}} = \langle \vec{s} \rangle_{\vec{f}_s}$.
 
 Given $N$ background images $\{\mathbf{b}_i\}_{i=1}^{N}$ and $N$ signal images $\{\mathbf{s}_i\}_{i=1}^{N}$, the empirical error to be minimized is:
  \begin{equation}\label{eq:L4}
  \begin{split}
   \hat{l}(\mathbf{w}) =& \frac{1}{N}\sum_{i=1}^{N}\Big\{[\mathbf{w}^T\mathbf{b}_i- \mathbf{w}^T\hat{\vec{b}}]^2 + 
   \frac{1}{2}[\mathbf{w}^T\mathbf{s}_i- \mathbf{w}^T\hat{\vec{s}}]^2\Big\}  \\
   + &\mathbf{{w}}^T\overline{\vec{K}}_\vec{n}\mathbf{{w}}  -2\mathbf{w}^T\hat{\vec{s}},
   \end{split}
  \end{equation}
  where $\hat{\vec{b}} = \frac{1}{N}\sum_{i=1}^{N}\mathbf{b}_i$, and $\hat{\vec{s}} = \frac{1}{N}\sum_{i=1}^{N}\mathbf{s}_i$.
  
  To approximate the Hotelling template, any gradient-based algorithm can be employed to minimize Eq. ($\ref{eq:L4})$. This method also does not require inversion of covariance matrix.

 \section{Numerical studies} \label{sec:num}
 Computer-simulation studies were conducted to investigate the proposed methods for learning the IO and HO test statistics.
Four different binary signal detection tasks were considered. A \textit{signal-known-exactly and background-known-exactly} (SKE/BKE) signal detection task was considered in which 
the IO and HO can be analytically determined. 
A \textit{signal-known-exactly and background-known-statistically} (SKE/BKS) detection task and a \textit{signal-known-statistically and background-known-statistically} (SKS/BKS) detection task that both employed a lumpy background object model~\cite{rolland1992effect} were also considered. 
For these two BKS signal detection tasks,
computations of the IO test statistic by use of MCMC methods have been accomplished~\cite{kupinski2003ideal, park2003ideal}. 
Finally, a SKE/BKS detection task employing a clustered lumpy background (CLB) object model~\cite{bochud1999statistical} was addressed. To the best of our knowledge, current MCMC applications to the CLB object model have not been reported~\cite{abbey2008ideal}. For all considered signal detection tasks, ROC curves were fit by use of the Metz-ROC software \cite{metz1998rockit} that utilized the ``proper'' binormal model~\cite{metz1999proper, pesce2007reliable}.

The imaging system in all studies was simulated by a linear C-D mapping with a Gaussian kernel that was motivated by an idealized parallel-hole collimator system~\cite{kupinski2003ideal, kupinski2003experimental}:
\begin{equation} \label{eq:kernel}
{h}_m(\vec{r}) = \frac{h}{2\pi w^2} \exp{\left(\frac{-\left(\vec{r}-\vec{r}_m\right)^{T}(\vec{r}-\vec{r}_m)}{2w^2}\right)},  
 \end{equation}
 where the height $h = 40$ and the width $w=0.5$. 
 The details for each signal detection task and the training of neural networks are given in the following subsections.
   
 \subsection{SKE/BKE signal detection task} \label{subsec:SKE}
 Both the signal and background were non-random for this case.
The image size was $64\times 64$ (i.e., $M=4096$) and the background image was specified as $\vec{b} = \vec{0}$.
 The signal function ${f}_s(\vec{r})$ was a 2D symmetric Gaussian function:
 \begin{equation}\label{eq:ske_s}
 {f}_s(\vec{r}) = A  \exp{\left(\frac{-\left(\vec{r}-\vec{r}_c\right)^{T}(\vec{r}-\vec{r}_c)}{2w_s^2}\right)},
 \end{equation} 
 where $A=0.2$ is the amplitude,  $\vec{r}_c = [32, 32]^T$ is the coordinate of the signal location, and $w_s = 3$ is the width of the signal. The {signal image $\vec{s}$} can be computed as:
 \begin{equation} \label{eq:SKE_s}
 {{s}}_m = \frac{Ahw_s^2}{(w^2+w_s^2)} \exp{\left(-\frac{(\vec{r}_m-\vec{r}_c)^{T}(\vec{r}_m-\vec{r}_c)}{2(w^2+w_s^2)}\right)}.
 \end{equation} 
 {Independent and identically distributed (i.i.d.) Laplacian noise that can describe histograms of filtered natural images~\cite{clarkson2000approximations} was employed: ${n}_m \sim \mathcal{L}(0, c)$}, where $\mathcal{L}(0, c)$ denotes a Laplacian distribution with the exponential decay $c$. The value of $c$ was set to $30 / \sqrt{2}$, which corresponds to standard deviation $30$. 
 
 Because the randomness in the measurements was only from the Laplacian noise, the IO test statistic can be computed as~\cite{clarkson2000approximations}:
 \begin{equation}\label{eq:BKE_IO}
 \begin{split}
  \Lambda_\text{LR}(\vec{g}) =\exp{\left[ \frac{1}{c}\sum_{m=1}^{M} (|{g}_m - {b}_m| - |{g}_m - {b}_m - {s}_m|) \right]}.
 \end{split}
 \end{equation}
The Hotelling template can be computed by analytically inverting the covariance matrix $\vec{K}_j\in \mathbb{R}^{M\times M}$ ($j=0, 1$):
\begin{equation}
\begin{split}
\vec{K}^{-1}_j(m,n) =
 \begin{cases}
	\frac{1}{2c^2}, &\text{if } m=n \\
	0,&\text{if } m\neq n, \\
\end{cases}
\end{split}
\end{equation}
where $\vec{K}^{-1}_j(m,n)$ denotes the component at the $m^{th}$ row and the $n^{th}$ column ($1\leq m,n \leq M$) of  $\vec{K}^{-1}_j$. 
The performances of the proposed learning-based methods were compared to those produced by these analytical computations for this case.

 \subsection{SKE/BKS signal detection task with a lumpy background model}\label{sec:skebke_lumpy}
 In this case, the image size was $64\times 64$ and a non-random signal described by Eq. (\ref{eq:ske_s}) was employed.
The background was random and described by a stochastic lumpy object model~\cite{rolland1992effect}:
 \begin{equation} \label{eq:lumpy}
{f}_{b}(\vec{r}) = \sum_{n=1}^{N_b}l(\vec{r}-\vec{r}_{n} | a, s),
\end{equation}
where $N_b$ is the number of lumps that is sampled from Poisson distribution with the mean $\overline{N}$: $N_b \sim \mathcal{P}(\overline{N})$, 
$\mathcal{P}(\overline{N})$ denotes a Poisson distribution with the mean $\overline{N}$ that was set to 5,
and $l(\vec{r}-\vec{r}_n | a, s)$ is the lumpy function modeled by a 2D Gaussian function with amplitude $a$ and width $s$: 
\begin{equation}
l(\vec{r}-\vec{r}_n | a, s) = a\exp{\left(-\frac{(\vec{r}-\vec{r}_n)^T(\vec{r}-\vec{r}_n)}{2s^2} \right)}. 
\end{equation}
Here, $a$ was set to 1, $s$ was set to 7, and $\vec{r}_{n}$ is the location of the $n^{th}$ lump that was sampled from uniform distribution over the field of view. 
The background image $\vec{b}$ was analytically computed as:
\begin{equation}
\begin{split}
  {{b}}_m =  \frac{ahs^2}{w^2+s^2}
  \sum_{n=1}^{N_b} \exp{\left(-\frac{(\vec{r}_n-\vec{r}_m)^{T}(\vec{r}_n-\vec{r}_m)}{2(w^2+s^2)}\right)}.
  \end{split}
\end{equation}
The measurement noise was an i.i.d. Gaussian noise that models electronic noise: ${n}_m \sim \mathcal{N}\left(0, \delta^2\right)$, where $\mathcal{N}\left(0, \delta^2\right)$ denotes a Gaussian distribution with the mean 0 and the standard deviation $\delta$ that was set to $20$.
Examples of signal-present images are shown in the top row of Fig. \ref{fig: CLB_imgs}.

The IO and HO test statistics cannot be analytically determined because of the background randomness.  To serve as a surrogate for ground truth, the MCMC method was employed to approximate the IO test statistic. In one Markov Chain, 200,000 background images were sampled according to the proposal density and the acceptance probability defined in~\cite{kupinski2003ideal}. 
The traditional HO test statistic was calculated by use of the covariance-matrix decomposition~\cite{barrett2013foundations} with an empirical background covariance matrix that was estimated by use of 100,000 background images. 
 
 \vspace{-0.5cm}
 \subsection{SKS/BKS signal detection task with a lumpy background model}
{This case employed the same stochastic lumpy background model that was specified in the SKE/BKS case described above.} The signal was random and modeled by a 2D Gaussian function with a random location and a random shape, which can be mathematically represented as:
\begin{equation} \label{eq:sks_signal}
\begin{split}
{f}_s&(\vec{r}) = A   
\exp{\left({-\left[\vec{R}_\theta\left(\vec{r}-\vec{r}_c\right)\right]^{T}\vec{D}^{-1}\left[\vec{R}_\theta(\vec{r}-\vec{r}_c)\right]}\right)}.
\end{split}
\end{equation} 
Here, $\vec{R}_\theta= \begin{bmatrix}
       \cos(\theta) & -\sin(\theta)           \\[0.3em]
       \sin(\theta) & \cos(\theta)
     \end{bmatrix}$
is {a rotation matrix} that rotates a vector through an angle $\theta$ in Euclidean space, and $\vec{D}=
\begin{bmatrix}
2w_1^2 & 0 \\
0 & 2w_2^2 
\end{bmatrix}
$ determines the width of the Gaussian function along each coordinate axis.  
The signal image $\vec{s}$ was analytically computed as:
 \begin{equation}
 {s}_m  = A'  \exp{\left({-\left[\vec{R}_\theta\left(\vec{r}-\vec{r}_c\right)\right]^{T}\vec{D}'^{-1}\left[\vec{R}_\theta(\vec{r}-\vec{r}_c)\right]}\right)},
 \end{equation} 
where $A' = {Ahw_1 w_2} \sqrt{\frac{1}{(w^2+w_1^2)(w^2+w_2^2)}}$ and $\vec{D}'= \begin{bmatrix}
2(w^2+w_1^2) & 0 \\
0 & 2(w^2+w_2^2) 
\end{bmatrix}$. The value of $A$ was set to $0.2$, $\theta$ was drawn from a uniform distribution: $\theta \sim \mathcal{U}(0,2\pi)$, $w_1$ and $w_2$ were sampled from a uniform distribution: $w_1, w_2 \sim \mathcal{U}(2, 4)$, and $\vec{r}_c$ was uniformly distributed over the image field of view. 
The measurement noise was Gaussian having zero mean and a standard deviation of $10$.

The MCMC method was employed to provide a surrogate for ground truth for the IO. In each Markov Chain, 400,000 background images were sampled according to the proposal density and the acceptance probability described in~\cite{park2003ideal}.
The traditional HO test statistic was calculated by use of the covariance-matrix decomposition~\cite{barrett2013foundations} with an empirical object covariance matrix that was estimated by use of 100,000 background images and 100,000 signal images. 

Because linear observers typically are unable to detect signals with random locations, the HO was expected to perform poorly.
Multi-template model observers~\cite{eckstein2001model, zhang2004automated, castella2009mass} and the scanning HO~\cite{barrett2006objective, gifford2005comparison} can be employed to detect variable signals.
In this paper, we do not provide a method for training these observers. {The approximation of multi-template observers and the scanning HO by use of a supervised learning method represents a topic for future investigation. }

 \subsection{SKE/BKS signal detection task with a clustered lumpy background model} \label{sec:CLB}
\begin{figure}[H]
\centering
\begin{subfigure}[b]{0.11\textwidth}
   \centering
 \includegraphics[width=1.0\linewidth]{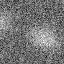}
  \vspace{-0.6cm}
 \caption{}
 \end{subfigure}
 \begin{subfigure}[b]{0.11\textwidth}
  \centering
 \includegraphics[width=1.0\linewidth]{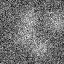}
  \vspace{-0.6cm}
 \caption{}
 \end{subfigure}
  \begin{subfigure}[b]{0.11\textwidth}
  \centering
 \includegraphics[width=1.0\linewidth]{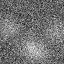}
  \vspace{-0.6cm}
 \caption{}
 \end{subfigure}
  \begin{subfigure}[b]{0.11\textwidth}
  \centering
 \includegraphics[width=1.0\linewidth]{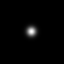}
  \vspace{-0.6cm}
 \caption{}
 \end{subfigure}

\vspace{0.3cm}
   \begin{subfigure}[b]{0.11\textwidth}
   \centering
 \includegraphics[width=1.0\linewidth]{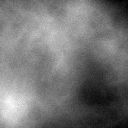}
 \vspace{-0.6cm}
 \caption{}
 \end{subfigure}
 \begin{subfigure}[b]{0.11\textwidth}
  \centering
 \includegraphics[width=1.0\linewidth]{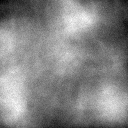}
  \vspace{-0.6cm}
 \caption{}
 \end{subfigure}
  \begin{subfigure}[b]{0.11\textwidth}
  \centering
 \includegraphics[width=1.0\linewidth]{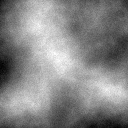}
  \vspace{-0.6cm}
 \caption{}
 \end{subfigure}
  \begin{subfigure}[b]{0.11\textwidth}
  \centering
 \includegraphics[width=1.0\linewidth]{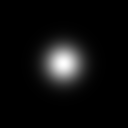}
  \vspace{-0.6cm}
 \caption{}
 \end{subfigure}
 \caption{(a)-(c) Samples of the signal-present measurements for the SKE/BKS detection task with the lumpy background model. (d) An image showing the signal contained in (a)-(c).   (e)-(g) Samples of the signal-present measurements for the SKE/BKS detection task with the CLB model. (h) An image showing the signal contained in (e)-(g).  }
 \label{fig: CLB_imgs}
\end{figure}

A second SKE/BKS detection task associated with a more sophisticated stochastic background model, the clustered lumpy background (CLB), was considered also. 
The CLB model can be employed
to synthesize mammographic images~\cite{bochud1999statistical}. In this study, the image size was set to $128\times128$ and a CLB realization was simulated as: 
 \begin{equation}
 {b}_m = \sum_{k=1}^{K}\sum_{n=1}^{N_k} l\left(\vec{r}_m - \vec{r}_k - \vec{r}_{kn}| \vec{R}_{\theta _{kn}} \right),
 \end{equation}
where $K\sim \mathcal{P}(\overline{{K}})$ is the number of clusters, $N_k\sim \mathcal{P}(\overline{{N}})$ is the number of blobs in the $k^{th}$ cluster, $\vec{r}_k$ is the location of the $k^{th}$ cluster, and $\vec{r}_{kn}$ is the location of the $n^{th}$ blob in the $k^{th}$ cluster. Here,  $\vec{r}_k$ was sampled from a uniform distribution over the image field of view, $\vec{r}_{kn}$ was sampled from a Gaussian distribution with standard deviation $\sigma$ and center $\vec{r}_k$, and $l(\vec{r} | \vec{R}_{\theta _{kn}} )$ is the blob function:
\vspace{-0.1cm}
\begin{equation}
l(\vec{r} | \vec{R}_{\theta_{kn}} ) = {a}\exp\left(-\alpha \frac{\|\vec{R}_{\theta_{kn}} \vec{r}\|^\beta}{\text{L}(\vec{R}_{\theta_{kn}}\vec{r})} \right),
\end{equation}
where $a$, $\alpha$ and $\beta$ are adjustable parameters. The rotation matrix $\vec{R}_{\theta_{kn}}$ is associated with the angle $\theta_{kn} \sim \mathcal{U}(0, 2\pi)$, and $\text{L}(\vec{r})$ is the ``radius" of the ellipse with half-axes $L_x$ and $L_y$:
\vspace{-0.25cm}
\begin{equation}
\text{L}(\vec{r}) =\frac { L_x L_y} {\sqrt {L_x^2 \sin^2 (\theta_\vec{r}) + L_y^2 \cos^2 (\theta_\vec{r}) }},
\end{equation}
where $\theta_\vec{r} = {\arctan (\frac{r_y}{r_x})}$. Here, $r_x$ and $r_y$ denote the components of $\vec{r}$.
The parameters employed for generating the CLB images are summarized in Table \ref{table}. 
\renewcommand{\arraystretch}{1.2}
\begin{table}[H]
\vspace{-0.17cm}
\centering
\caption{Parameters for generating CLB images.}
\vspace{-0.16cm}
\begin{tabular}{| c | c | c | c | c | c | c | c |}
\hline
{$\overline{\text{K}}$} & $\overline{\text{N}}$ & Lx & Ly & $\alpha$ & $\beta$ & $\sigma$ & $a$\\ \hline
                       150  & 20  & 5  & 2   & 2.1  & 0.5 & 12 & 100   \\ \hline
\end{tabular}
\label{table}
\end{table}
The signal image was generated as a 2D symmetric Gaussian function centered in the image with an amplitude of $500$ and a width of $12$. Mixed Poisson-Gaussian noise that models both photon noise and electronic noise was employed. The standard deviation of Gaussian noise was set to $10$. 
Examples of signal-present images are shown in the bottom row of Fig.~\ref{fig: CLB_imgs}.

To the best of our knowledge,  current MCMC methods have not been applied to the CLB object model and the mixed Poisson-Gaussian noise model.
To provide a surrogate for ground truth for the HO,
the traditional HO was computed by use of covariance-matrix decomposition with the empirical background covariance matrix estimated using 400,000 background images. 

\subsection{Details of training neural networks}
Here, details regarding the implementation of the supervised learning-based methods for approximating the IO and HO
for the tasks above are described.

The train-validation-test scheme~\cite{goodfellow2016deep} was employed to evaluate the proposed supervised learning approaches.
Specifically, the CNNs and SLNNs were trained on a training dataset. Subsequently, these neural networks were specified based upon a validation dataset and the detection performances of these networks were finally assessed on a testing dataset.
To prepare training datasets for the BKS detection tasks, 100,000 lumpy background~\cite{rolland1992effect} images and 400,000 CLB images~\cite{bochud1999statistical} were generated. 
When training the CNNs for approximating IOs, to mitigate the overfitting that can be caused by insufficient training data, 
a ``semi-online learning" method was proposed and employed. In this approach,
the measurement noise was generated on-the-fly and added to noiseless images drawn from the finite datasets.
The validation dataset and testing dataset both comprised 200 images for each class.

 To approximate the HO test statistic, SLNNs that represent linear functions were trained by use of the proposed method employing the covariance-matrix decomposition described in Sec. \ref{sec:HO_cmd}. This was possible because the noise models for the considered detection tasks were known. 
 At each iteration in training processes, the parameters of SLNNs were updated by minimizing error function Eq. (\ref{eq:L4}) on mini-batches drawn from the training dataset. 
 Specifically, when training the SLNN for the SKE/BKE detection task,
 the signal and background that were known exactly were employed and each mini-batch contained the fixed signal image and background image.
When training the SLNNs for the SKE/BKS detection tasks, the known signals were employed and each mini-batch contained 200 background images and the fixed signal image.
For training the SLNN for the SKS/BKS detection task,
 each mini-batch contained 200 background images and 200 signal images. 
 The weight vector $\vec{w}$ that produced the maximum $\text{SNR}_t$ value evaluated on the validation dataset was specified to approximate the Hotelling template. The feasibility of the proposed methods for approximating the HO from a reduced number of images was also investigated.
 Specifically,
the SLNNs were trained for the SKE/BKS detection task with the CLB model by minimizing Eq. (\ref{eq:L3}) and Eq. (\ref{eq:L4}) on datasets comprising 2000 labeled measurements (contained 1000 signal-present images and 1000 signal-absent images) and 2000 background images, respectively.
 
\begin{figure}[t]
\includegraphics[width=\linewidth]{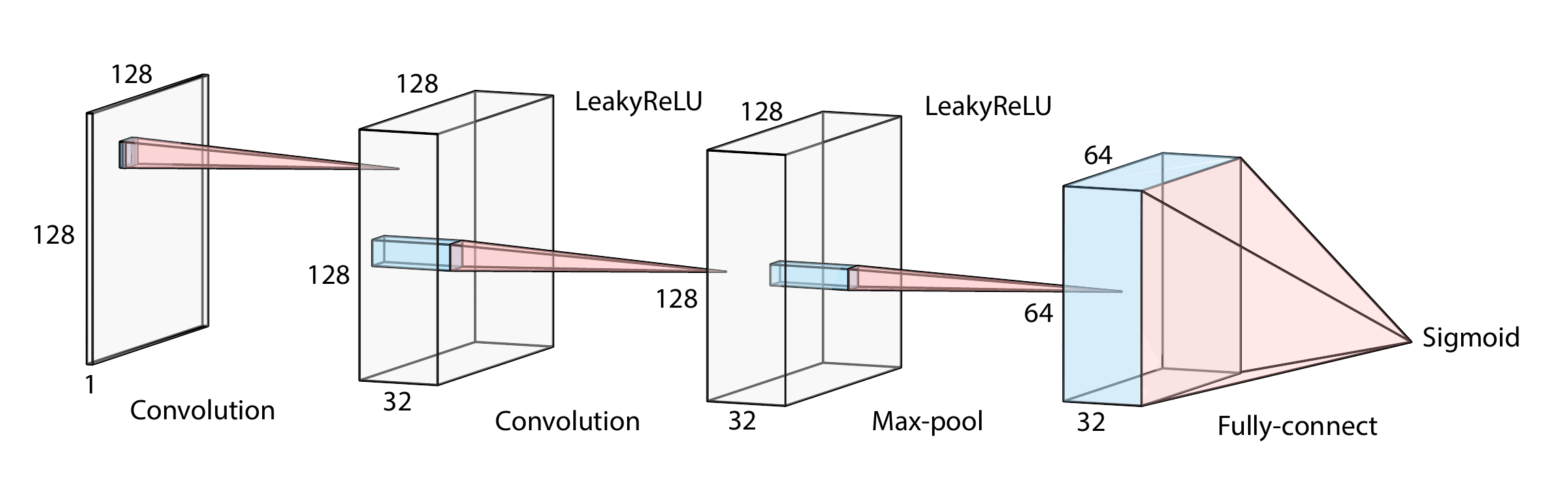}
 \caption{One instance of the CNN architecture employed for approximating the IO test statistic.}
 \label{fig:CNN_arc}
 \end{figure}

 As opposed to the case of the HO approximation where the network architecture is known linear,
to specify the CNN architecture for approximating the IO, 
a family of CNNs that possess different numbers of convolutional (CONV) layers was explored.
Specifically,
an initial CNN having one CONV layer was firstly trained by minimizing the cross-entropy described in Eq. (\ref{eq:IO_loss}). 
Subsequently, CNNs having additional CONV layers were trained according to Eq. (\ref{eq:IO_loss}) until the 
network did not significantly decrease the cross-entropy on a validation dataset. 
The cross-entropy was considered as significantly decreased if its decrement is at least $1.0\%$ of that produced by the previous CNN. 
Finally the CNN having the minimum validation cross-entropy was selected as the optimal CNN in the explored architecture family. 
For all the considered CNN architectures in this architecture family, each CONV layer comprised 32 filters with $5\times 5$ spatial support
and was followed by a LeakyReLU activation function~\cite{springenberg2014striving},
a max-pooling layer~\cite{scherer2010evaluation} following the last CONV layer was employed to subsample the feature maps, and finally
a fully connected (FC) layer using a sigmoid activation function computed the posterior probability $\Pr(H_1|\vec{g},\bm{\Theta})$. 
It should be noted that these architecture parameters were determined heuristically and may not be optimal for many signal detection tasks. 
One instance of the implemented CNN architecture is illustrated in Fig. \ref{fig:CNN_arc}.
These CNNs were trained by minimizing the error function defined in Eq. (\ref{eq:IO_loss}) on mini-batches at each iteration. Each mini-batch contained 200 signal-absent images and 200 signal-present images.
Because the HO detection performance is a lower bound of the IO detection performance,
the selected optimal CNN should not perform worse than the SLNN-approximated HO (SLNN-HO) on the corresponding signal detection task if that CNN approximates IO.
If this occurs, the architecture parameters need to be re-specified and a different family of CNN architectures should be considered.
 
 The Adam algorithm~\cite{kingma2014adam}, which is a stochastic gradient descent algorithm, was employed in Tensorflow~\cite{abadi2016tensorflow} to minimize the error functions for approximating the IO and HO. All networks were trained on a single NVIDIA TITAN X GPU.

\section{Results} \label{sec:results}
\subsection{SKE/BKE signal detection task}

\subsubsection{HO approximation}
\begin{figure}[H]
\centering
\includegraphics[width=0.91\linewidth]{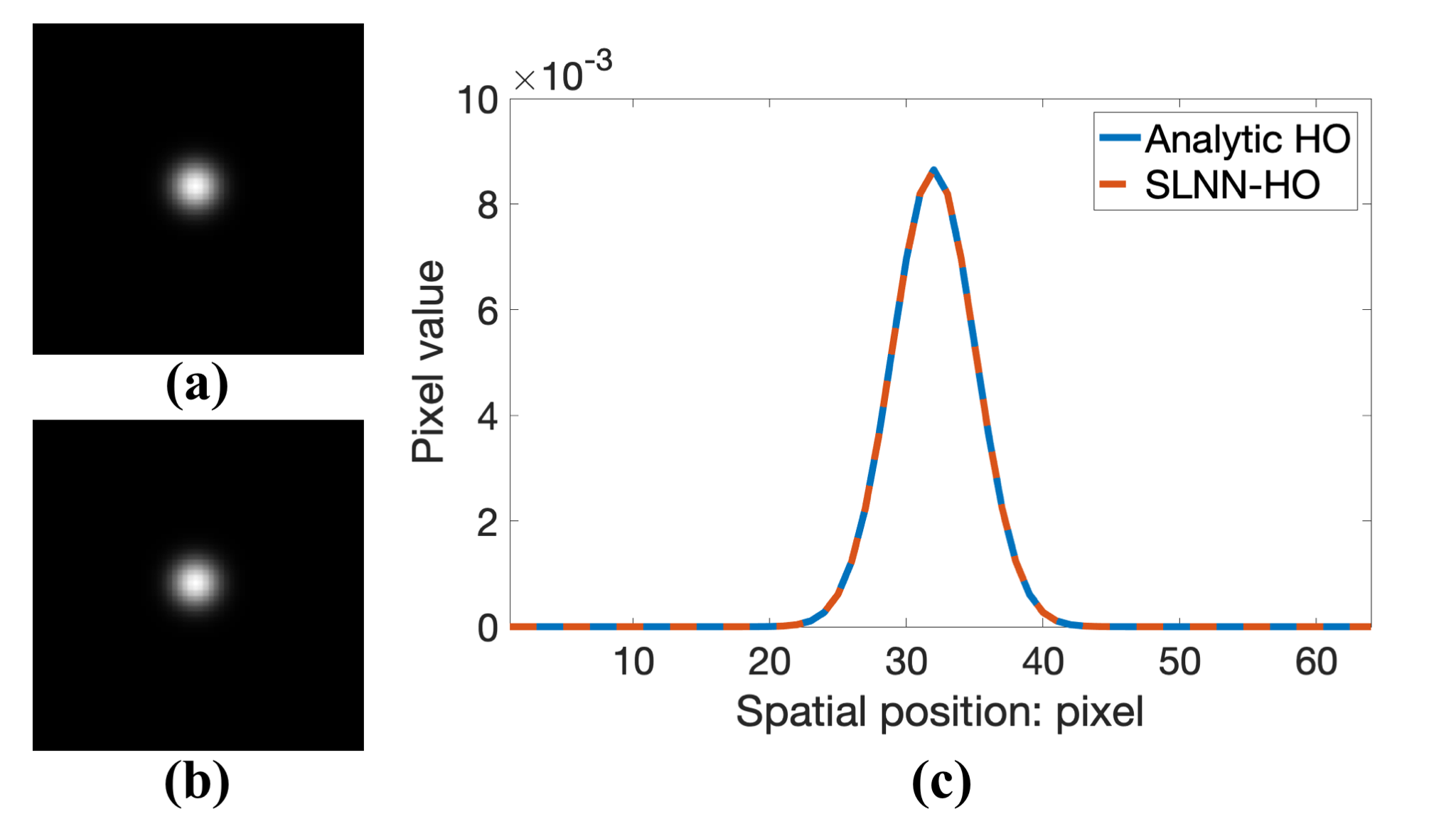}
\caption{Comparison of the Hotelling template in the SKE/BKE case: (a) Analytical Hotelling template; (b) SLNN-HO template; (c) Center line profiles in (a) and (b). The estimated templates are nearly identical.}
\label{fig:BKE_HO}
\end{figure}

 A linear SLNN was trained for 1000 mini-batches and 
 the weight vector $\vec{w}$ that produced the maximum $\text{SNR}_t$ value evaluated on the validation dataset was selected to approximate the Hotelling template.
The linear templates employed by the SLNN-HO and the analytical HO are shown in Fig. \ref{fig:BKE_HO}.
The results corresponding to the SLNN-HO closely approximate those of the analytical HO.
The ROC curve produced by the SLNN-HO (purple dashed curve) is compared to that produced by the analytical HO (yellow curve) in Fig. \ref{fig:BKE_ROC} (b). These two curves nearly overlap.

\subsubsection{IO approximation}
\begin{figure}[H]
\centering
 \begin{subfigure}{0.24\textwidth}
{\includegraphics[width=1.0\linewidth]{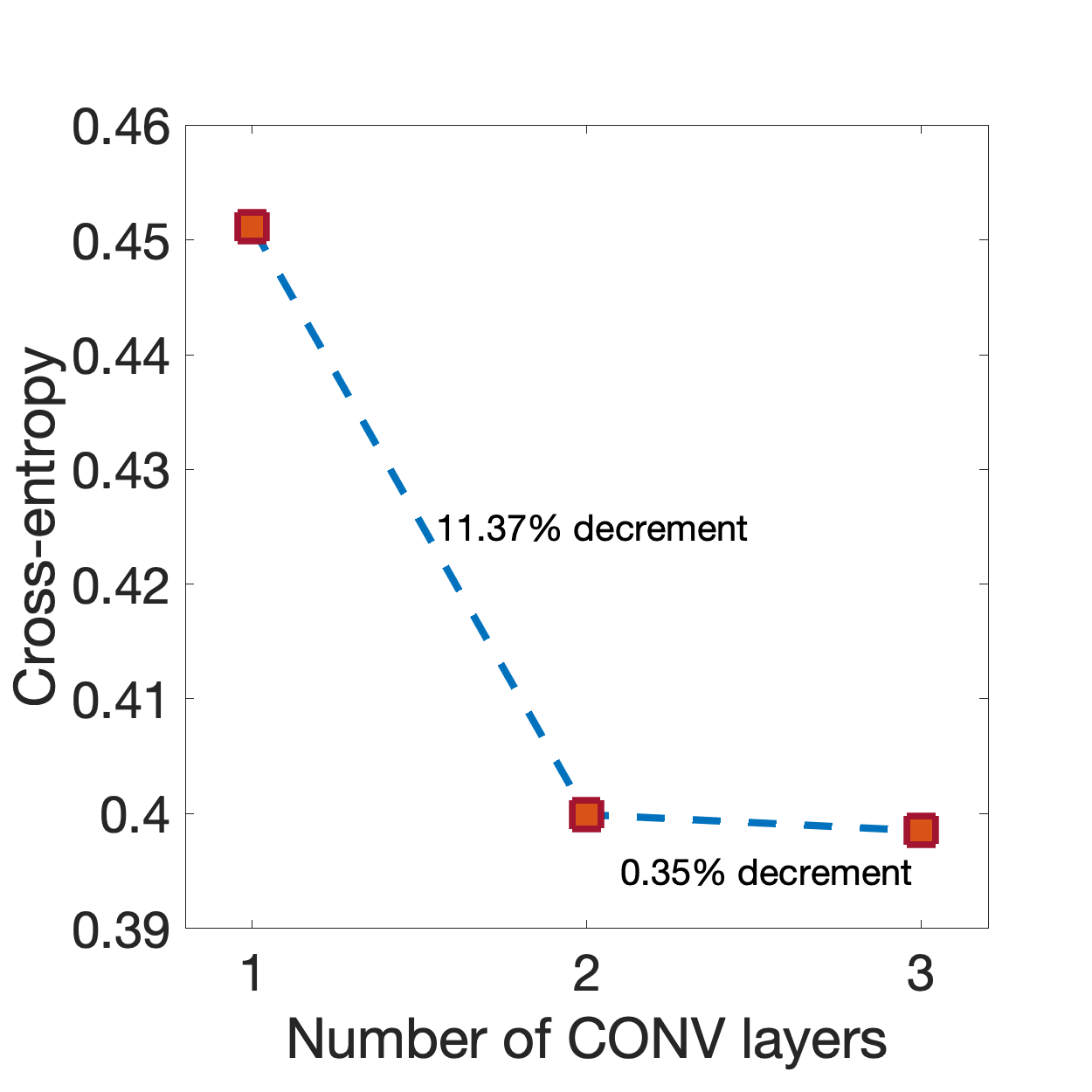}}
   \vspace{-0.6cm}
 \caption{}
 \end{subfigure}
  \begin{subfigure}{0.24\textwidth}
{\includegraphics[width=1.0\linewidth]{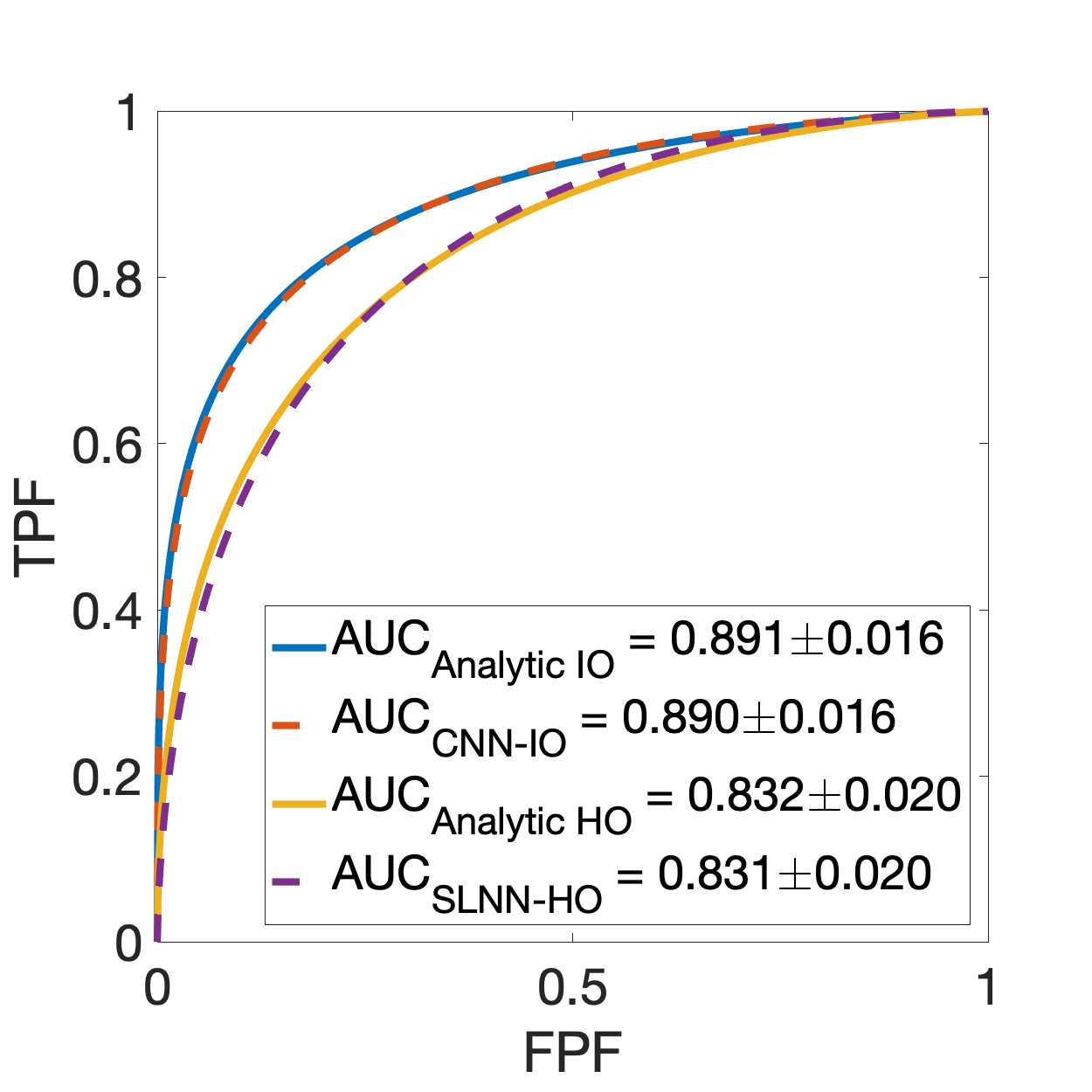}}
   \vspace{-0.6cm}
 \caption{}
 \end{subfigure}
\caption{ (a) Validation cross-entropy values of CNNs having one to three CONV layers; (b) Testing ROC curves for the IO and HO approximations.}
\label{fig:BKE_ROC}
\end{figure}

The CNNs having one to three CONV layers were trained for 100,000 mini-batches and the corresponding validation cross-entropy values are plotted in
Fig. \ref{fig:BKE_ROC} (a).  
The validation cross-entropy was not significantly decreased after adding the third CONV layer. Therefore, 
we stopped adding more CONV layers and the CNN having the minimum validation cross-entropy, which was the CNN that possesses 3 CONV layers, was selected.
The detection performance of this selected CNN was evaluated on the testing dataset and the resulting AUC value was 0.890, which was greater than that of the SLNN-HO (i.e., 0.831).
Subsequently, the selected CNN was employed to approximate the IO.
The testing ROC curve of the CNN-approximated IO (CNN-IO) (red-dashed curve) was compared to that of the analytical IO (blue curve) in Fig. \ref{fig:BKE_ROC} (b). 
The efficiency of the CNN-IO, which can be computed as the squared ratio of the detectability index~\cite{park2005efficiency} of the CNN-IO to that of the IO, was $99.14\%$. The mean squared error (MSE) of the posterior probabilities computed by the analytical IO and the CNN-IO was $0.30\%$. These quantities were evaluated on the testing dataset.

\subsection{SKE/BKS signal detection task with lumpy background}
 \subsubsection{HO approximation}
\begin{figure}[H]
\centering
\includegraphics[width=0.91\linewidth]{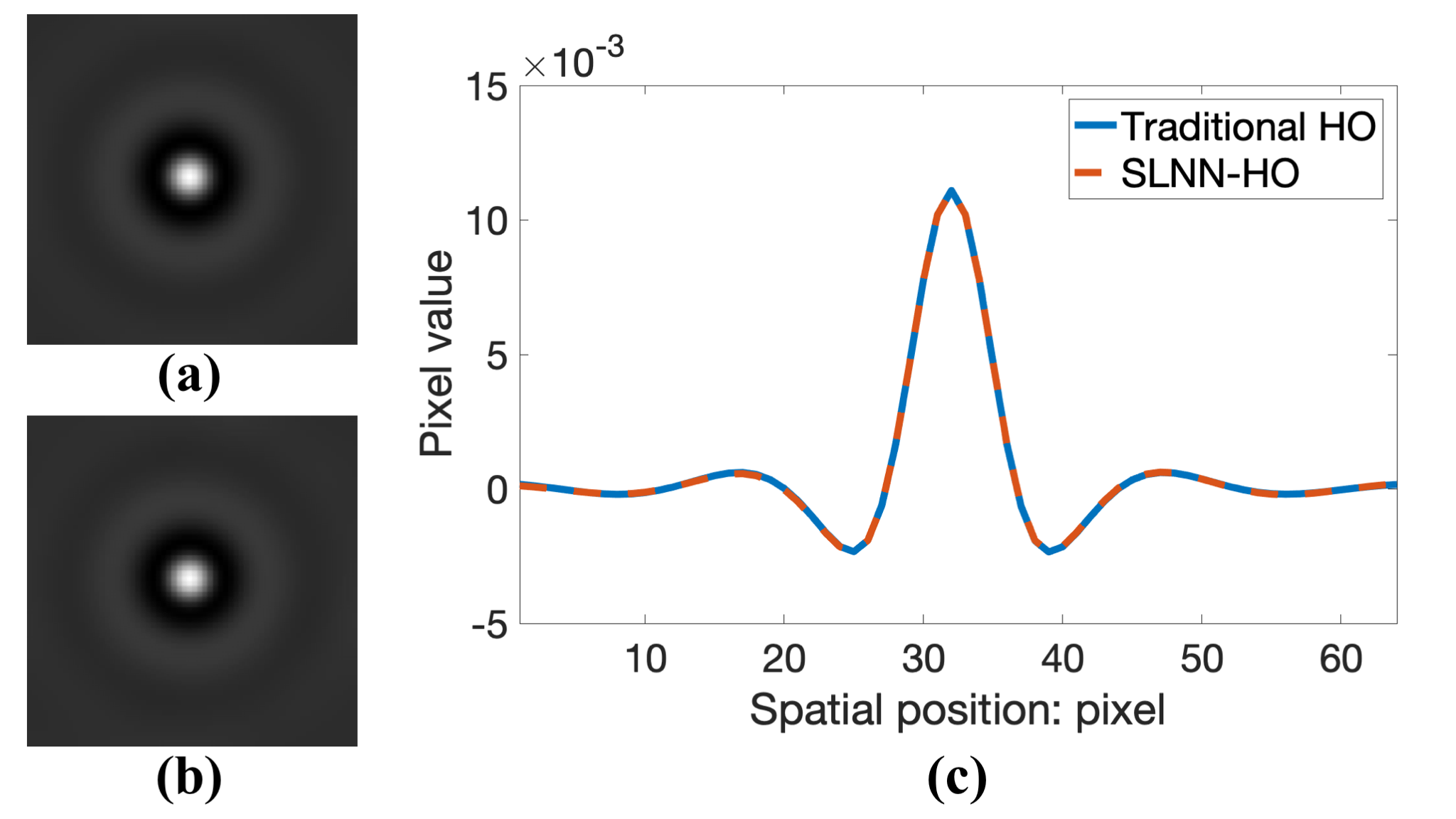}
\caption{Comparison of the Hotelling template in the SKE/BKS case: (a) Traditional Hotelling template; (b) SLNN-HO template; (c) Center line profiles in (a) and (b). The estimated templates are nearly identical.}
\label{fig:BKS_HO}
\end{figure}

 The SLNN was trained for 1000 mini-batches (i.e., 2 epochs) and 
 the weight vector $\vec{w}$ that produced the maximum $\text{SNR}_t$ value evaluated on the validation dataset was selected to approximate the Hotelling template.
The linear templates employed by the SLNN-HO and the traditional HO are shown in Fig. \ref{fig:BKS_HO}. 
The results corresponding to the SLNN-HO closely approximate those of the traditional HO.
The ROC curves corresponding to the traditional HO (yellow curve) and the SLNN-HO (purple-dashed curve) are compared in Fig. \ref{fig:BKS} (b). Two ROC curves nearly overlap.

\subsubsection{IO approximation}
\begin{figure}[H]
\centering
 \begin{subfigure}{0.24\textwidth}
 \includegraphics[width=1.0\linewidth]{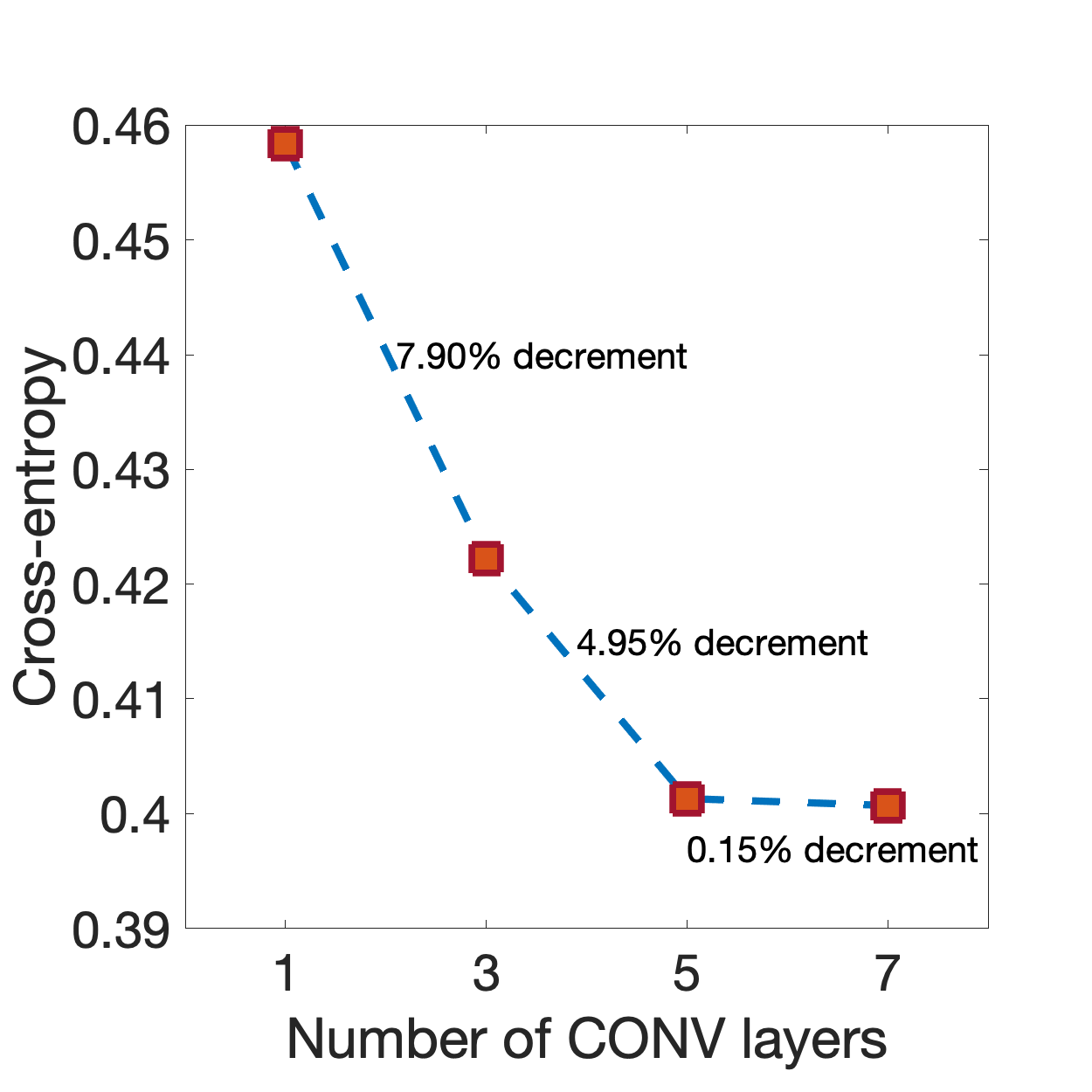}
  \vspace{-0.6cm}
 \caption{}
 \end{subfigure}
  \begin{subfigure}{0.24\textwidth}
 \includegraphics[width=1.0\linewidth]{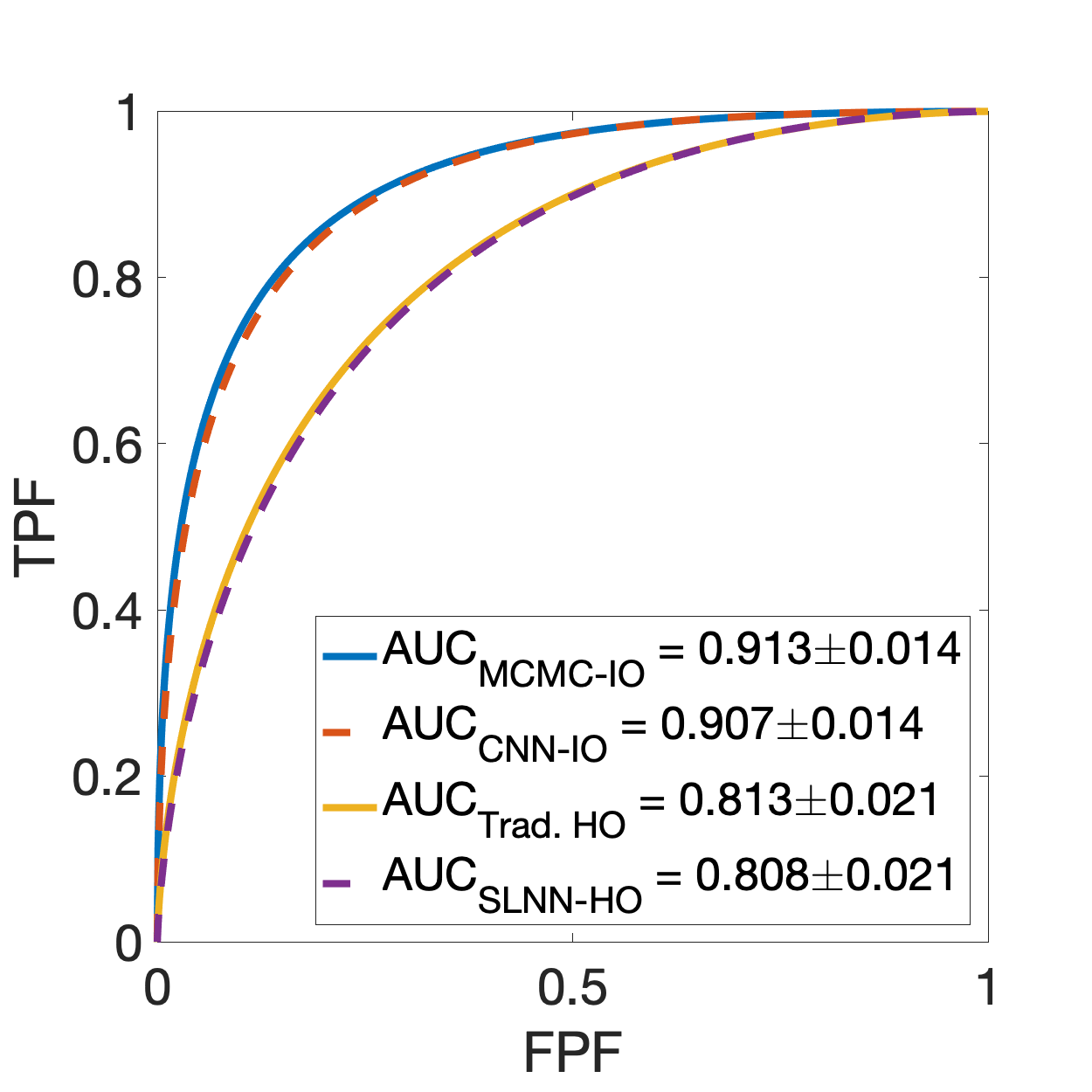}
  \vspace{-0.6cm}
 \caption{}
 \end{subfigure}
\caption{(a) Validation cross-entropy values of CNNs having one to seven CONV layers; (b) Testing ROC curves for the IO and HO approximations.}
\label{fig:BKS}
\end{figure}

The CNNs having 1, 3, 5, and 7 CONV layers were trained for 100,000 mini-batches (i.e., 200 epochs) and the corresponding validation cross-entropy values are plotted in
Fig. \ref{fig:BKS} (a).  
There was no significant difference of the validation cross-entropy between the CNNs having 5 and 7 CONV layers. Therefore, 
we stopped adding more CONV layers and the CNN having the minimum validation cross-entropy, which was the CNN that possesses 7 CONV layers, was selected.
The selected CNN was evaluated on the testing dataset and the resulting AUC value was 0.907, which was greater than that of the SLNN-HO (i.e., 0.808).
Subsequently, the selected CNN was employed to approximate the IO.
The testing ROC curve of the CNN-IO (red-dashed curve) is compared to that of the MCMC-computed IO (MCMC-IO) (blue curve) in Fig. \ref{fig:BKS} (b). The efficiency of the CNN-IO was $94.64\%$ with respect to the MCMC-IO, and the MSE of the posterior probabilities computed by the CNN-IO and the MCMC-IO was $0.84\%$. These quantities were evaluated on the testing dataset.

\subsection{SKS/BKS signal detection task with lumpy background}
\subsubsection{HO approximation}
\begin{figure}[H]
\centering
\includegraphics[width=0.91\linewidth]{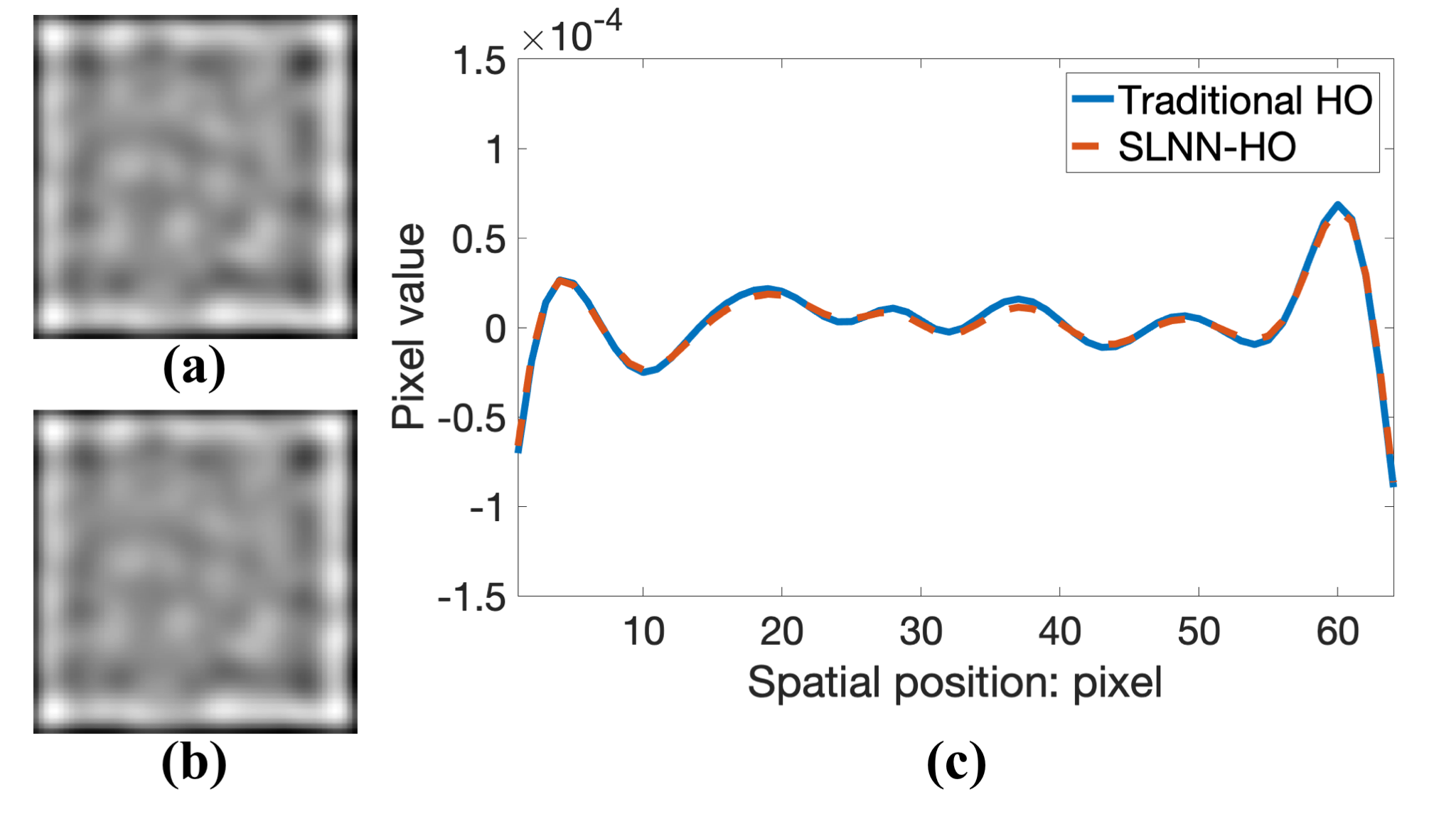}
\caption{Comparison of the Hotelling template in the SKS/BKS case: (a) Traditional Hotelling template; (b) SLNN-HO template; (c) Center line profiles in (a) and (b). The estimated templates are nearly identical.}
\label{fig:SKS_HO}
\end{figure}

 A linear SLNN was trained for 1000 mini-batches (i.e., 2 epochs) and 
 the weight vector $\vec{w}$ that produced the maximum $\text{SNR}_t$ value evaluated on the validation dataset was selected to approximate the Hotelling template.
The linear templates employed by the SLNN-HO and the traditional HO are shown in Fig. \ref{fig:SKS_HO}.
The results corresponding to the SLNN-HO closely approximate those of the traditional HO.
The ROC curves corresponding to the SLNN-HO (purple dashed curve) and the traditional HO (yellow curve) are compared in Fig. \ref{fig:SKS_IO} (b). 
The two ROC curves nearly overlap. The HO performed nearly as a random guess for this task as expected.

\subsubsection{IO approximation}
\begin{figure}[H]
\centering
 \begin{subfigure}{0.24\textwidth}
 \includegraphics[width=1.0\linewidth]{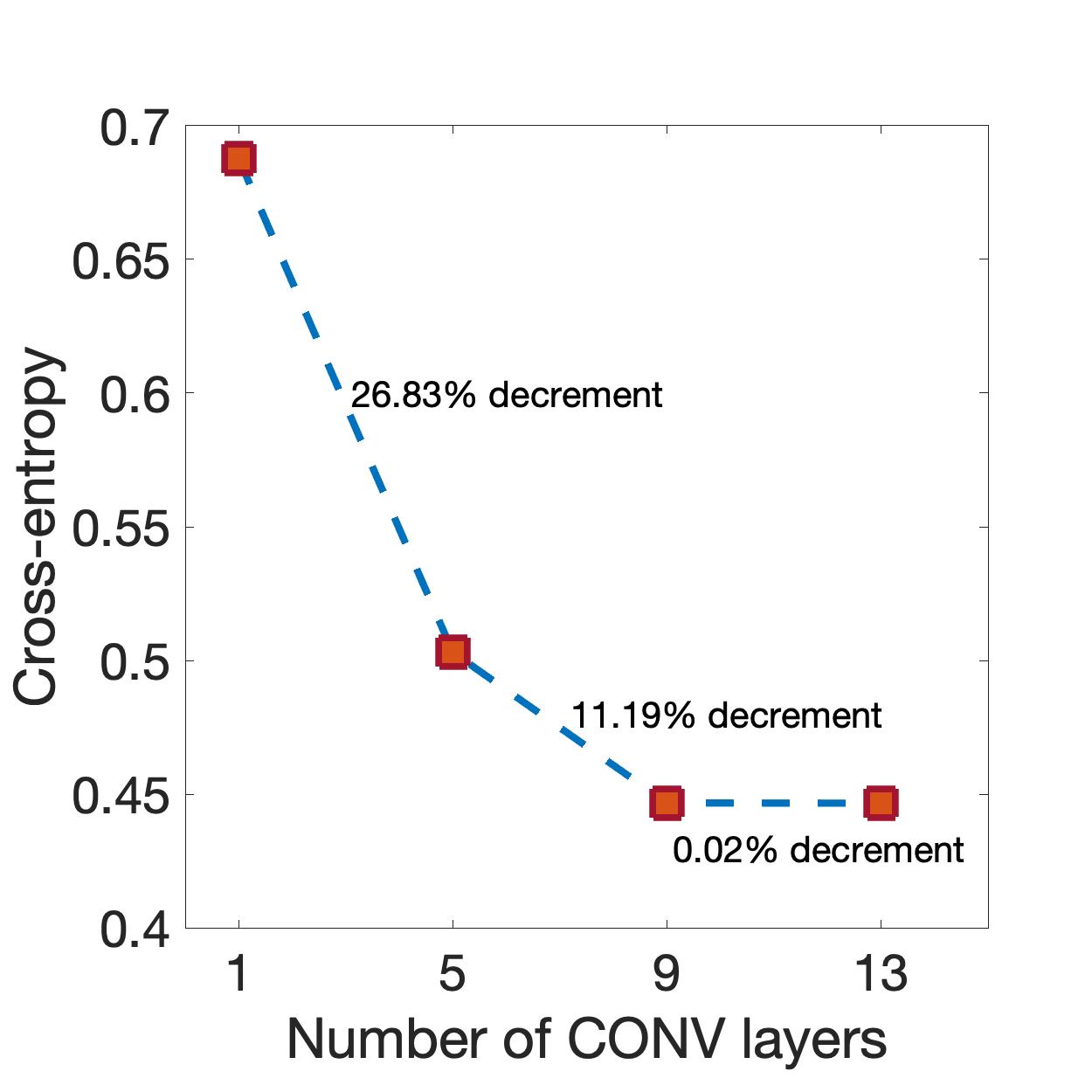}
 \vspace{-0.6cm}
 \caption{}
 \end{subfigure}
  \begin{subfigure}{0.24\textwidth}
 \includegraphics[width=1.0\linewidth]{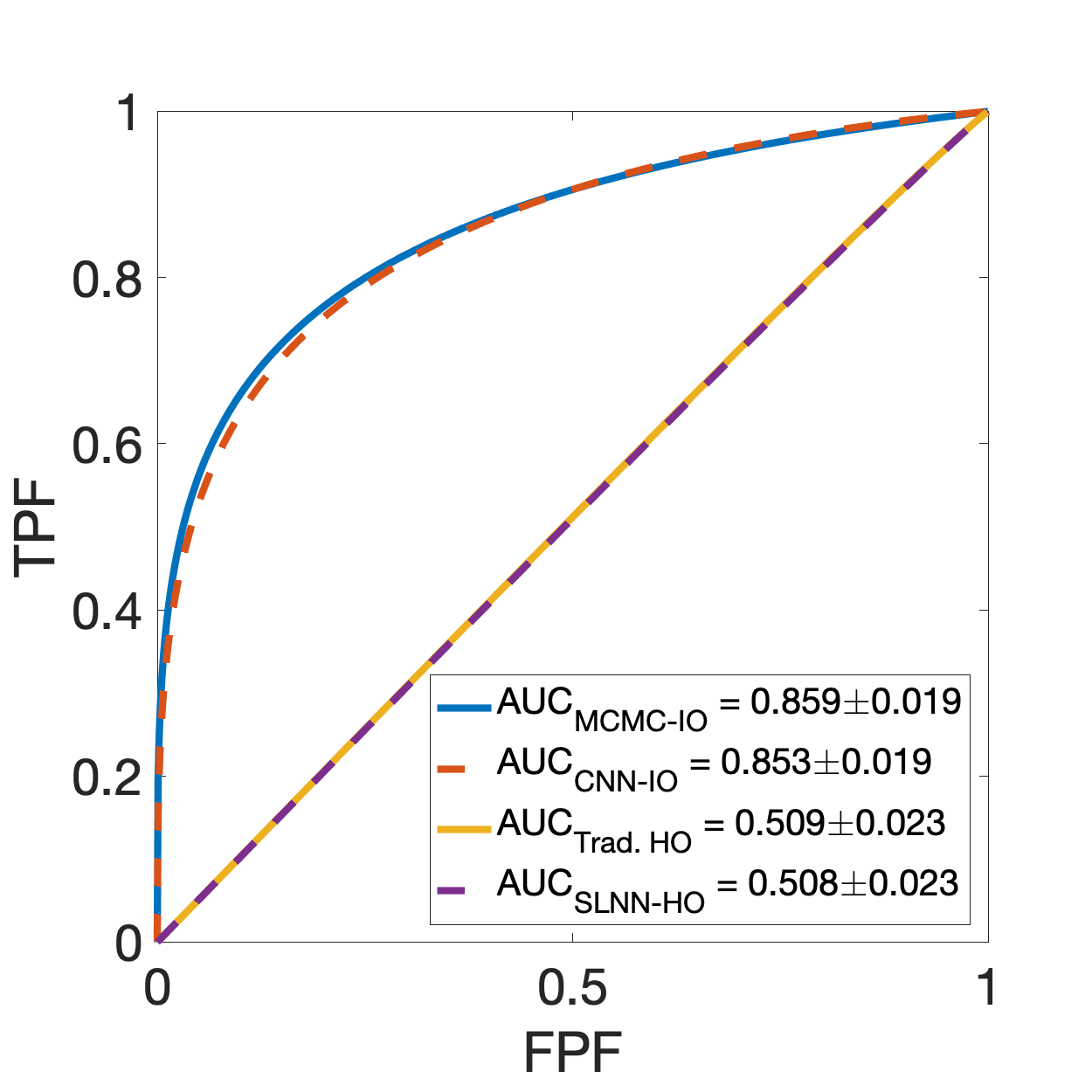}
 \vspace{-0.6cm}
 \caption{}
 \end{subfigure}
\caption{(a) Validation cross-entropy values produced by CNNs having 1 to 13 CONV layers; (b) Testing ROC curves for the IO and HO approximations.}
\label{fig:SKS_IO}
\end{figure}

Convolutional neural networks having 1, 5, 9, and 13 CONV layers were trained for 300,000 mini-batches~(i.e., 600 epochs) and the corresponding validation cross-entropy values are plotted in Fig. \ref{fig:SKS_IO} (a).
Because there was no significant decrement of the validation cross-entropy value after adding 4 CONV layers to the CNN having 9 CONV layers, we stopped adding more CONV layers and the CNN having the minimum validation cross-entropy value, which was the CNN with 13 CONV layers, was selected.
The selected CNN was evaluated on the testing dataset and the resulting AUC value was 0.853, which was greater than that of the SLNN-HO (i.e., 0.508).
Subsequently, the selected CNN was employed to approximate the IO.
 The testing ROC curve produced by the CNN-IO (red-dashed curve) is compared to that produced by the MCMC-IO (blue curve) in Fig. \ref{fig:SKS_IO} (b). The efficiency of the CNN-IO was $95.14\%$ with respect to the MCMC-IO and the MSE of the posterior probabilities computed by the CNN-IO and the MCMC-IO was $1.46\%$. These quantities were evaluated on the testing dataset.

\subsubsection{CNN visualization}
\begin{figure}[H]
\centering
 \begin{subfigure}[b]{0.125\textwidth}
  \centering
 \includegraphics[width=1.0\linewidth]{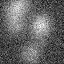}
   \vspace{-0.6cm}
 \caption{}
 \end{subfigure}
  \begin{subfigure}[b]{0.125\textwidth}
  \centering
 \includegraphics[width=1.0\linewidth]{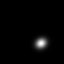}
   \vspace{-0.6cm}
 \caption{}
 \end{subfigure}
  \begin{subfigure}[b]{0.125\textwidth}
  \centering
 \includegraphics[width=1.0\linewidth]{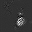}
   \vspace{-0.6cm}
 \caption{}
 \end{subfigure}
 
   \vspace{0.03cm}
  \begin{subfigure}[b]{0.125\textwidth}
  \centering
 \includegraphics[width=1.0\linewidth]{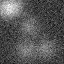}
   \vspace{-0.6cm}
 \caption{}
 \end{subfigure}
  \begin{subfigure}[b]{0.125\textwidth}
  \centering
 \includegraphics[width=1.0\linewidth]{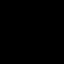}
   \vspace{-0.6cm}
 \caption{}
 \end{subfigure}
  \begin{subfigure}[b]{0.125\textwidth}
  \centering
 \includegraphics[width=1.0\linewidth]{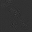}
   \vspace{-0.6cm}
 \caption{}
 \end{subfigure}
 \caption{(a) Signal-present measurements; (b) Image showing the signal contained in (a); (c) The signal feature map corresponding to (a); (d) Signal-absent measurements; (e) Image showing that the signal is absent in (d); (f) The signal feature map corresponding to (d). In the signal feature maps, the regions around the signals were activated by the CNN.}
  \label{fig:act_map}
 \end{figure}

 Feature maps extracted by CONV layers enabled us to understand how CNNs were able to extract task-specific features for performing signal detection tasks.
In this case, the 32 subsampled feature maps output from the max-pooling layer were weighted by the weight parameters of the last FC layer and then summed to produce a single 2D image for the visualization. 
That single 2D image was referred to as the signal feature map and is shown in Fig. \ref{fig:act_map}. 
The signal to be detected was nearly invisible in the signal-present measurements but can be easily observed in the signal feature map. This illustrates the ability of CNNs to perform signal detection tasks. 
 
 \subsection{SKE/BKS signal detection task with clustered lumpy background}
\subsubsection{HO approximation}
\begin{figure}[H]
\centering
\includegraphics[width=1.0\linewidth]{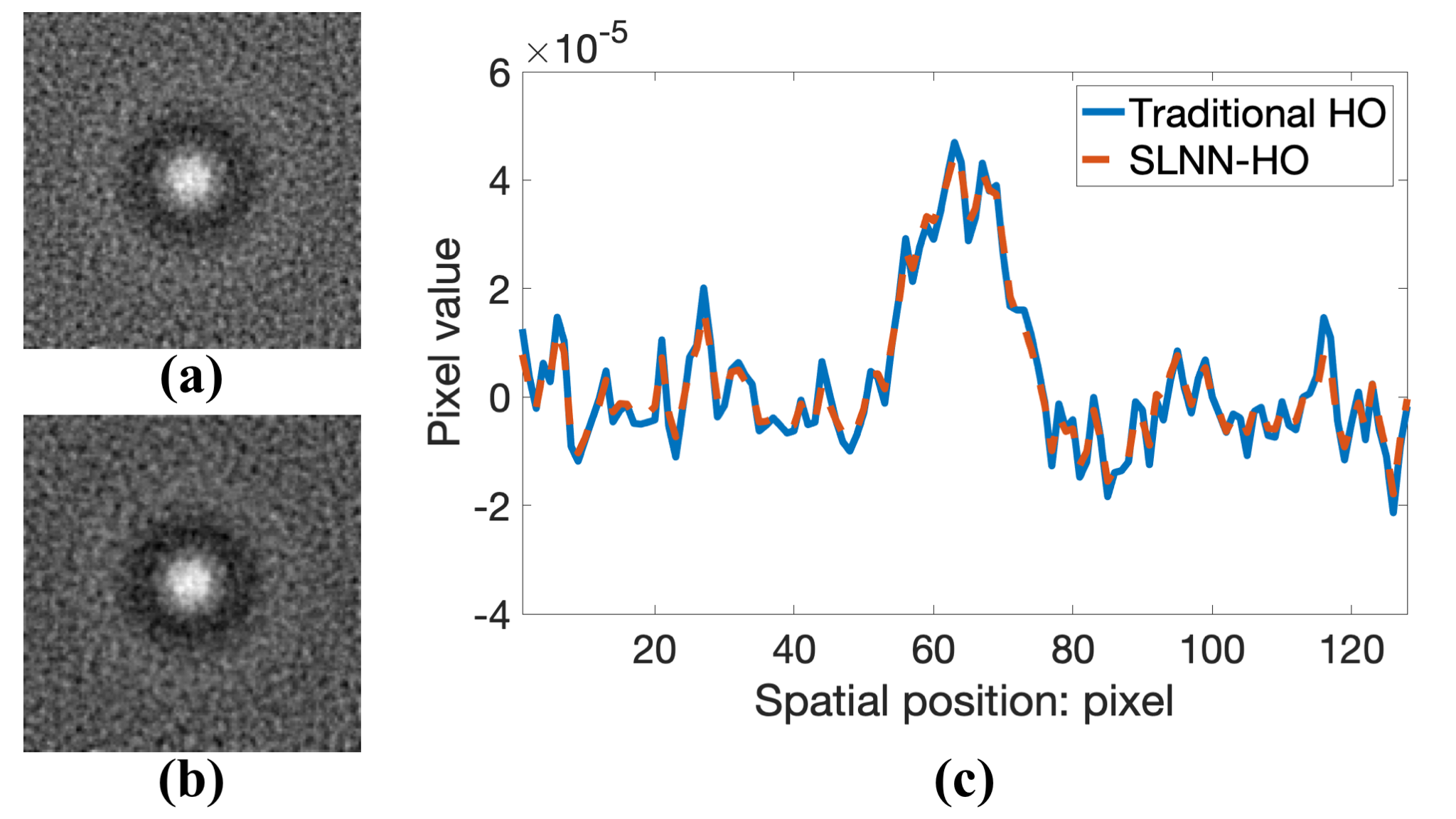}
\vspace{-0.6cm}
\caption{Comparison of the Hotelling template: (a) Traditional Hotelling template; (b) SLNN-HO template; (c) Center line profiles in (a) and (b). The estimated templates are nearly identical.}
\label{fig:CLB_HO}
\end{figure}

\begin{figure}[H]
\centering
 \begin{subfigure}{0.24\textwidth}
 \includegraphics[width=1.0\linewidth]{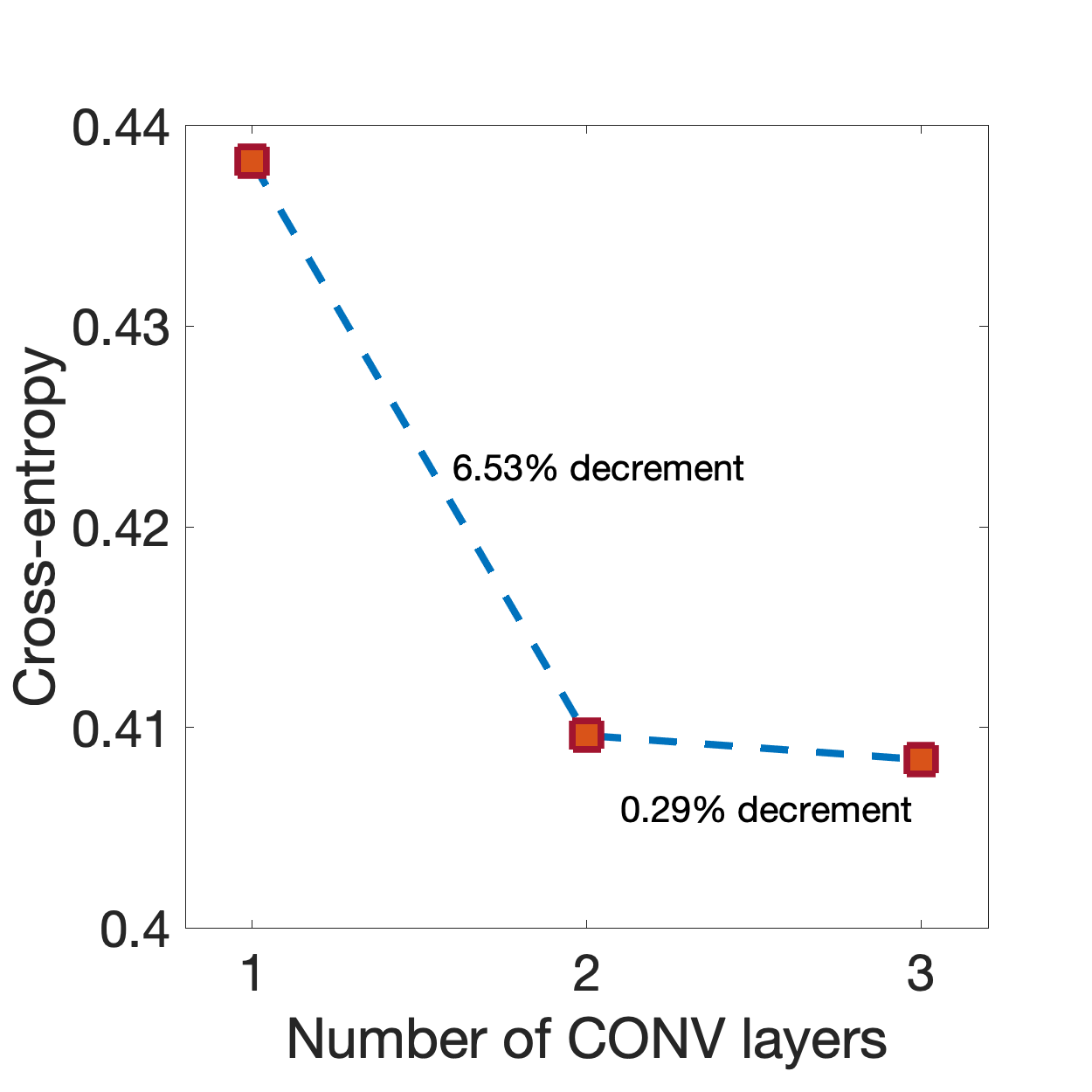}
  \vspace{-0.6cm}
 \caption{}
 \end{subfigure}
  \begin{subfigure}{0.24\textwidth}
 \includegraphics[width=1.0\linewidth]{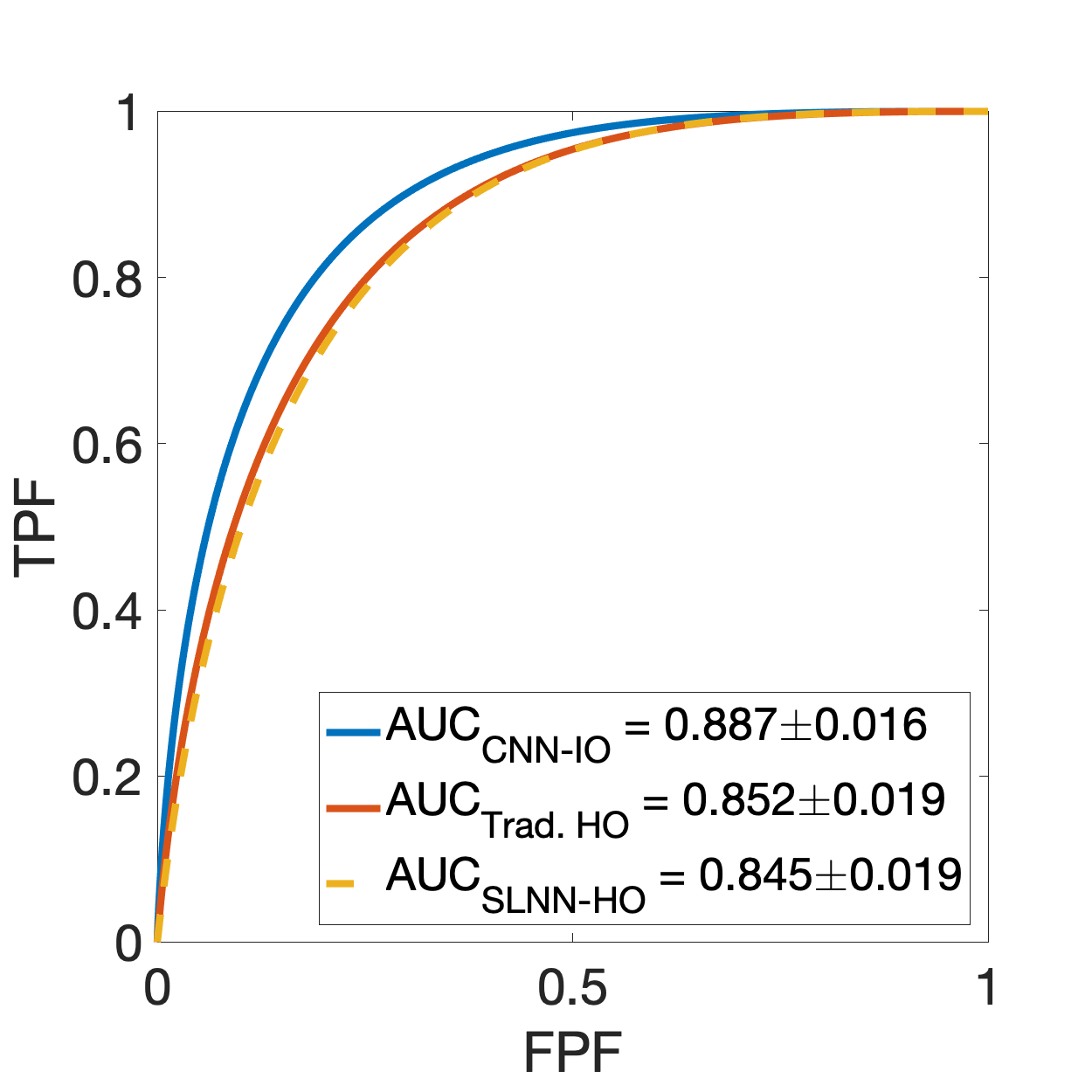}
  \vspace{-0.6cm}
 \caption{}
 \end{subfigure}
 \vspace{-0.2cm}
\caption{(a) Validation cross-entropy values of CNNs having one to three CONV layers; (b) Testing ROC curves for the IO and HO approximations. }
\label{fig:CLB}
\end{figure}

 The SLNN was trained for 40,000 mini-batches (i.e., 20 epochs) and 
 the weight vector $\vec{w}$ that produced the maximum validation $\text{SNR}_t$ was selected to approximate the Hotelling template.
The traditional HO template and the SLNN-HO template are compared in Fig. \ref{fig:CLB_HO}. 
The results corresponding to the SLNN-HO closely approximate those of the traditional HO.
\vspace{-0.3cm}
The ROC curve of the SLNN-HO (yellow-dashed curve) compares to that of the traditional HO (red curve) in Fig. \ref{fig:CLB} (b). Two curves nearly overlap.
  \vspace{-0.4cm}

\vspace{-0.3cm}
 \subsubsection{IO approximation}
\begin{figure}[H]
\centering
\includegraphics[width=0.9\linewidth]{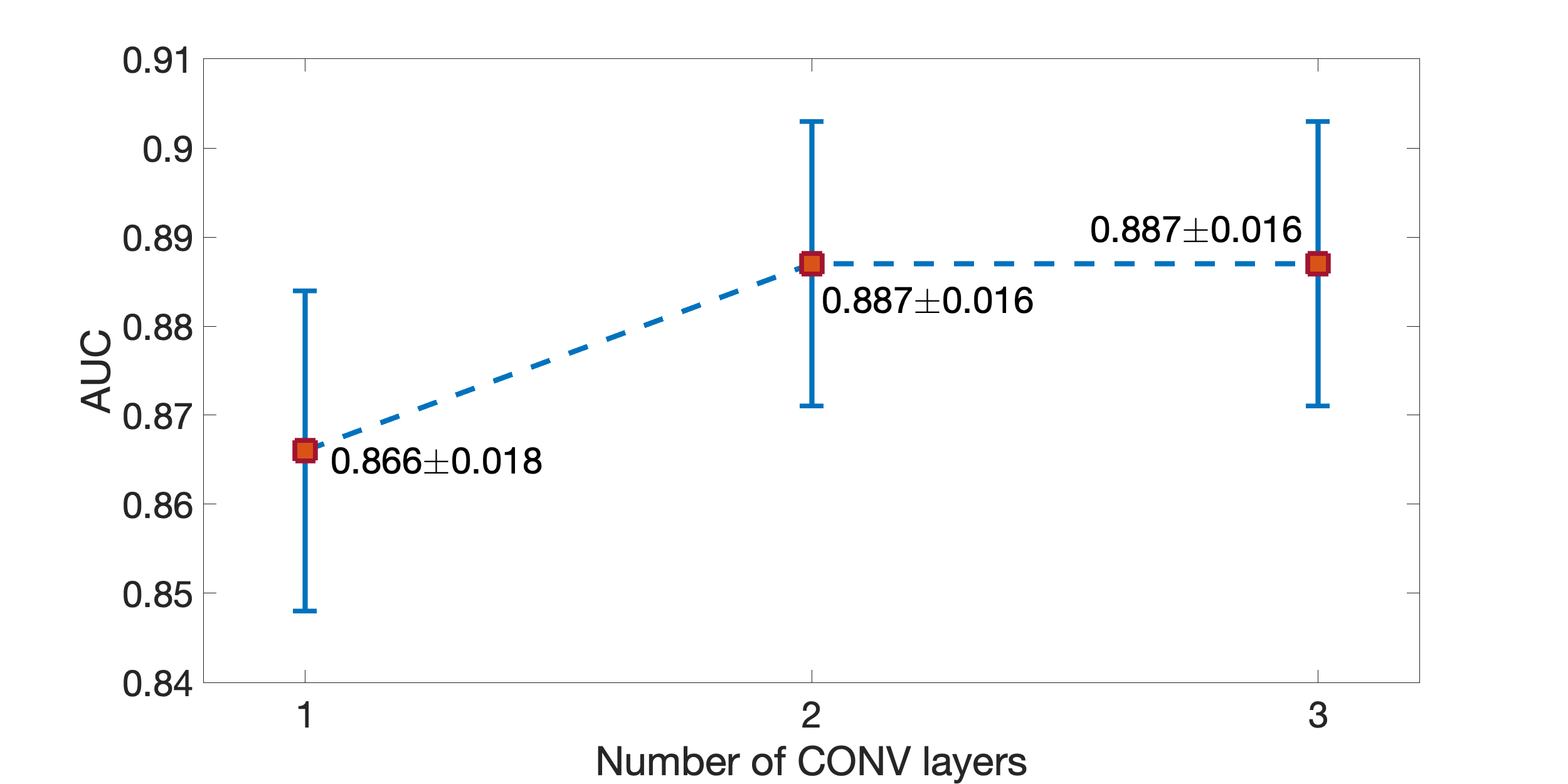}
\vspace{-0.1cm}
\caption{Testing AUC values of CNNs having one to three CONV layers.}
\label{fig:CLB_AUCs}
\end{figure}

 Convolutional neural networks having one to three CONV layers were trained for 100,000 mini-batches (i.e., 50 epochs) and the corresponding validation cross-entropy values are plotted in Fig. \ref{fig:CLB} (a). Because the validation cross-entropy was not significantly decreased by adding the third CONV layer,
we stopped adding more CONV layers and 
the CNN having the minimum validation cross-entropy value, which was the CNN with three CONV layers, was selected.
The detection performance of this selected CNN was evaluated on the testing dataset and the resulting AUC value was 0.887, which was greater than that of the SLNN-HO (i.e., 0.845).
Subsequently, the selected CNN was employed to approximate the IO.
The CNN-IO was evaluated on the testing dataset and the resulting ROC curve is plotted in Fig.~\ref{fig:CLB} (b).
To show how the signal detection performance varied when the number of CONV layers was increased, the AUC values evaluated on the testing dataset corresponding to the CNNs with one to three CONV layers are illustrated in Fig. \ref{fig:CLB_AUCs}. These AUC values were estimated by use of the ``proper'' binormal model~\cite{metz1999proper, pesce2007reliable}. The AUC value was increased when more CONV layers were employed until convergence.
  \vspace{-0.4cm}
\vspace{-0.3cm}
Because MCMC applications to the CLB object model have not been reported to date, validation for the IO approximation was not provided in this case.
To the best of our knowledge, we are the first to approximate the IO test statistic for the CLB object model.

\subsubsection{HO approximation from a reduced number of images}
\begin{figure}[H]
\centering
 \begin{subfigure}{0.24\textwidth}
 \includegraphics[width=1.0\linewidth]{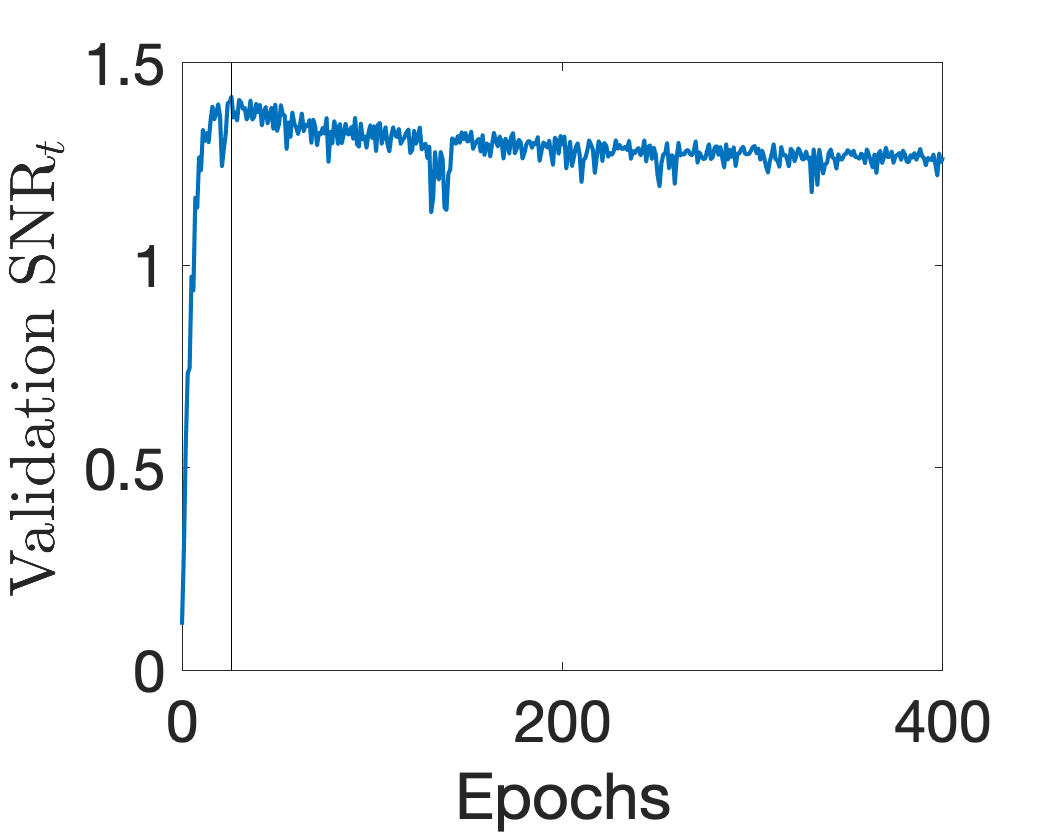}
  \vspace{-0.6cm}
 \caption{}
 \end{subfigure}
  \begin{subfigure}{0.24\textwidth}
 \includegraphics[width=1.0\linewidth]{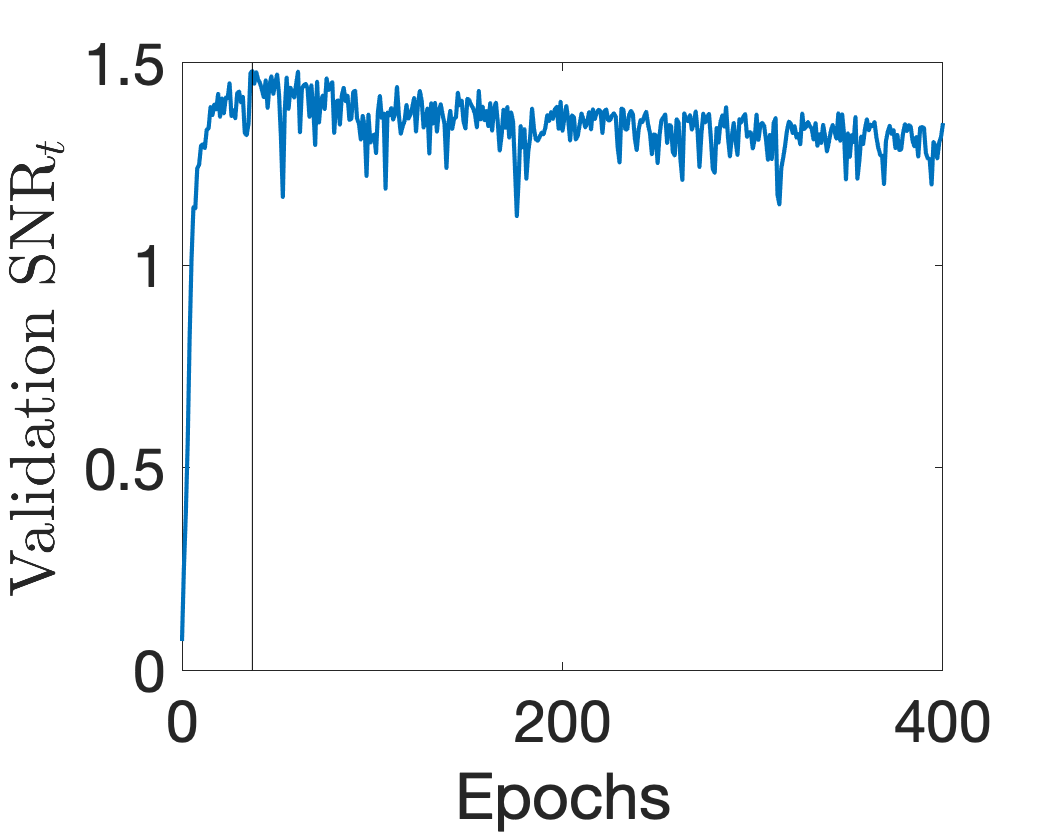}
  \vspace{-0.6cm}
 \caption{}
 \end{subfigure}
 \vspace{-0.2cm}
\caption{Curves of validation $\text{SNR}_t$ with respect to the number of epochs. (a) Validation $\text{SNR}_t$ curve of the SLNN trained on labeled noisy measurements. (b) Validation $\text{SNR}_t$ curve of the SLNN trained on background images using decomposition of covariance matrix. The vertical gray line indicates the epoch having the maximum validation $\text{SNR}_t$ value. Overfitting occurred after the overall curves of validation $\text{SNR}_t$ start to decrease.}
\label{fig:finite_curve}
\end{figure}

To solve the dimensionality problem of inverting a large covariance matrix for computing the Hotelling template, the matrix-inversion lemma has been implemented in which the covariance matrix is approximated by use of a
small number of images~\cite{barrett2013foundations}. 
However,
this method can introduce significant positive bias on the estimate of $\text{SNR}_{HO}$~\cite{kupinski2007bias}. 
To investigate the ability of our proposed methods to approximate the HO performance when small dataset is employed,
the linear SLNNs were trained by minimizing Eq. (\ref{eq:L3}) and Eq. (\ref{eq:L4}) on 2000 noisy measurements and 2000 background images, respectively, for 400 epochs. 
In the training processes, overfitting occurred as revealed by the curves of validation $\text{SNR}_t$ with respect to the number of epochs shown in Fig. \ref{fig:finite_curve}.
  \vspace{-0.3cm}
\vspace{-0.3cm}
However, an early-stopping strategy can be employed in which training is stopped at the epoch having the maximum validation  $\text{SNR}_t$.
The values of $\text{SNR}_{HO}^2$, which were computed according to Eq. (\ref{eq:SNR_HO}), evaluated at the $400^{th}$ epoch and at the epoch having the maximum validation $\text{SNR}_t$ are shown in Table \ref{table:SNR}. These data reveal that overfitting caused a significant positive bias on $\text{SNR}_{HO}^2$ while the early-stopping strategy accurately approximated the reference $\text{SNR}_{HO}^2$, which was computed by using the Hotelling template of the traditional HO that was shown in Fig. \ref{fig:CLB_HO} (a).
The Hotelling template was also computed by using the matrix-inversion lemma~\cite{barrett2013foundations} on 2000 background images, and the corresponding $\text{SNR}_{HO}^2$ had a significant positive bias shown in Table \ref{table:SNR} as observed by others~\cite{kupinski2007bias}.  
  \vspace{-0.2cm}
\begin{table}[H]
\centering
\caption{$\text{SNR}_{HO}^2$ computed from both background images $\vec{b}$ and measurements $\vec{g}$. The Hotelling template computed from few images can cause significant positive bias. However, when SLNNs were trained using our proposed methods, early-stopping strategy in which the epoch having the maximum validation $\text{SNR}_t$ was selected could be employed to closely approximate the HO performance.}
\vspace{-0.1cm}
\resizebox{\columnwidth}{!}{
\begin{tabular}{| c | c | c | c | c | c | c | c |}
\hline
Methods   & $400^{th}$ epoch & Early-stopping \\ \hline
Minimizing Eq. (\ref{eq:L3})      &     4.0421      &             \bf{2.0940}                  \\ \hline
Minimizing Eq. (\ref{eq:L4})    &       3.1101    &                \bf{2.1380}            \\ \hline
Matrix-inversion lemma       & \multicolumn{2}{c|}{5.7979}                   \\ \hline
Reference & \multicolumn{2}{c|}{\bf{2.1075} }                   \\ \hline
\end{tabular}
}
\label{table:SNR}
\end{table}

\vspace{-0.53cm}
\section{Discussion and Conclusion} \label{sec:concludes}
\vspace{-0.06cm}
The proposed supervised learning-based method that employs CNNs to approximate the IO test statistic represents an alternative approach to conventional numerical approaches such as MCMC methods for use in optimizing medical imaging systems and data-acquisition designs.
Although theoretical convergence properties exist for MCMC methods,  
practical issues such as designs of proposal densities from which proposed object samples are drawn need to be addressed for each considered object model and current applications of the MCMC methods have been limited to some specific object models that include parameterized torso phantoms~\cite{he2008toward}, lumpy background models~\cite{kupinski2003ideal} and a binary texture model~\cite{abbey2008ideal}.
Supervised learning-based approaches may be easier to deploy with sophisticated object models than are MCMC methods.  
To demonstrate this, in the numerical study, we applied the proposed supervised learning method with a CLB object model, for which the IO computation has not been addressed by MCMC methods to date~\cite{abbey2008ideal}. A practical advantage of the proposed method is that supervised learning-based methods are becoming widespread in their usage and
many researchers are becoming experienced on training feed-forward ANNs.

A challenge in approximating the IO by use of CNNs is the specification of the collection of model architectures to be systematically explored.
In this study,
we explored a family of CNNs that possess different numbers of CONV layers. By adding more CONV layers, the representation capacity of the network is increased and the test statistic can be more accurately approximated. This study does not provide methods for determining other architecture parameters such as the number of FC layers and the size of convolutional filters.  Recent work~\cite{cortes2016adanet} proposed a method that optimizes the network architecture in the training process. This represents a possible approach for jointly optimizing the network architecture and weights to approximate the IO test statistic.

We also proposed a supervised learning-based method using a simple linear SLNN
to approximate the HO that is the optimal linear observer and sets a lower bound of the IO performance. The proposed methodology directly learns the Hotelling template without estimating and inverting covariance matrices. Accordingly, the proposed method can scale well to large images. 
When approximating the HO test statistic, selection of network architecture is not an issue because the HO test statistic depends linearly on the input image and one can employ a linear SLNN to represent linear functions. 
We also provided an alternative method to learn the HO by use of a covariance-matrix decomposition. 
The feasibility of both methods to learn the HO from a reduced number of images was investigated.
For the case where 2000 clustered lumpy images with the dimension $128\times128$ were employed to approximate the HO, our proposed learning-based methods could still produce accurate estimates of $\text{SNR}_{HO}$ by incorporating an early-stopping strategy.

Numerous topics remain for future investigation. With regards to approximating IOs by use of experimental images, there is a need to investigate methods to train large CNN models on limited training data. To accomplish this, one may investigate transfer learning~\cite{qiu2016survey} or domain adaptation methods~\cite{ganin2014unsupervised} that learn features of images in target domain~(e.g., experimental images) by use of images in source domain~(e.g., computer-simulated images). 
One may also employ the method proposed by Kupinski~\emph{et al.}~\cite{kupinski2003experimental} or train a generative adversarial network~\cite{goodfellow2014generative} to estimate a stochastic object model (SOM) from experimental images to produce large datasets.
Finally, it will be important to extend the proposed learning-based methods to more complicated tasks, such as joint detection and localization of a signal. 

\vspace{-0.2cm}
\section*{Acknowledgment}
\vspace{-0.05cm}
This research was supported in part by NIH awards
EB020168 and EB020604 and NSF award DMS1614305.
\vspace{-0.17cm}






%
\bibliography{CNN_new}{}
\bibliographystyle{IEEEtran}


%
%
%
%
%




\end{document}